\documentclass[sigconf,screen]{acmart}

 \usepackage{flushend}
\usepackage{setspace}
\usepackage{array}
\usepackage{amsfonts}           
\usepackage{mathrsfs}
\usepackage{comment}
\usepackage{xargs}
\usepackage{mathtools}
\usepackage{tabularx}
\usepackage{enumerate}
\usepackage{amsthm}
\usepackage{enumitem}
\setlist{nosep}
\usepackage{tikz}
\usepackage[htt]{hyphenat}
\usepackage{color}
\usepackage{lscape}
\usepackage{blindtext}
\usepackage{hyperref}
\hypersetup{
    breaklinks=true,
    bookmarksopen,
    bookmarksdepth=2,
    breaklinks=true
}
\usepackage[T1]{fontenc}
\usepackage{capt-of}
\usepackage{url}
\usepackage{xurl}
\usepackage[framemethod=TikZ]{mdframed}
\usepackage{caption} 

\usepackage[normalem]{ulem}

\usepackage{amsmath}
\usepackage{graphicx}
\usepackage{multirow}
\usepackage{subfigure}
\usepackage{footnote}
\usepackage{listings}
\usepackage{xspace}
\usepackage{booktabs}
\usepackage[ruled,vlined,linesnumbered,noend]{algorithm2e}
\SetKwComment{Comment}{$\triangleright$\ }{}
\SetKwInOut{Variables}{Variables}
\usepackage{soul}
\usepackage{algorithmicx}  
\usepackage{algpseudocode}  
\usepackage{pifont}
\usepackage[flushleft]{threeparttable}
\usepackage{xcolor}
\usepackage{makecell}
\usepackage{lipsum}
\usepackage{balance}
\usepackage{caption} 
\usepackage{enumitem}
\setlist{noitemsep}

\definecolor{dkgreen}{rgb}{0,0.6,0}
\definecolor{gray}{rgb}{0.5,0.5,0.5}
\definecolor{mauve}{rgb}{0.58,0,0.82}
\definecolor{applegreen}{rgb}{0.55, 0.71, 0.0}
\definecolor{amber}{rgb}{1.0, 0.75, 0.0}
\definecolor{firebrick}{rgb}{0.7, 0.13, 0.13}
\definecolor{darkblue}{rgb}{0,0,0.55}

\usepackage{siunitx}
\newcommand{\pie}[1]{%
\begin{tikzpicture}
 \draw (0ex,0ex) circle (1ex);
 \fill (0ex,-1ex) arc (-90:(#1-90):1ex) -- (0ex,-1ex) -- cycle;
\end{tikzpicture}%
}

\def\eg{\emph{e.g.,}\xspace}

\def\ie{\emph{i.e.,}\xspace}
\def\etal{\emph{et al.}\xspace}
\def\vs{\emph{vs.}\xspace}

\newcommand{\one}{({\em i}\/)\xspace}
\newcommand{\two}{({\em ii}\/)\xspace}
\newcommand{\three}{({\em iii}\/)\xspace}

\usepackage[framemethod=TikZ]{mdframed}
\setlist{noitemsep}

\newcommand{\new}[1]{\textcolor{black}{#1}}

\newcommand{\detectionRateDecreaseAvg}[0]{91.75\%\xspace}
\newcommand{\detectionRateDecreaseDrebinPercAvg}[0]{74.11\%\xspace}
\newcommand{\detectionRateDecreaseMaMaPercAvg}[0]{52.50\%\xspace}
\newcommand{\detectionRateMaMaPercAvg}[0]{36.54\%\xspace}
\newcommand{\detectionRateDecreaseAvgWinPE}[0]{35.59\%\xspace}

\newcommand{\bypassVTEngines}[0]{37.35\xspace}
\newcommand{\bypassVTEnginesPerc}[0]{62.25\%\xspace}
\newcommand{\vtScoreOri}[0]{32.04\xspace}
\newcommand{\vtScoreAdv}[0]{22.65\xspace}
\newcommand{\vtScoreOriPerc}[0]{53.40\%\xspace}
\newcommand{\vtScoreAdvPerc}[0]{37.75\%\xspace}

\DeclareMathOperator{\performance}{DetectionRate}
\DeclareMathOperator{\vectorize}{vectorize}
\DeclareMathOperator{\shap}{shap}

\DeclareMathOperator{\map}{map}
\DeclareMathOperator{\manipulate}{manipulateFeature}

\DeclareMathOperator{\Gen}{Gen}
\DeclareMathOperator{\getindex}{getIndex}
\DeclareMathOperator{\argmin}{arg\,min}
\DeclareMathOperator{\argmax}{arg\,max}
\DeclareMathOperator{\range}{getRange}
\DeclareMathOperator{\countlarge}{countLarge}
\DeclareMathOperator{\ismanipulable}{isManipulatable}

\setcopyright{acmlicensed}
\acmPrice{15.00}
\acmDOI{10.1145/3611643.3616309}
\acmYear{2023}
\copyrightyear{2023}
\acmSubmissionID{fse23main-p595-p}
\acmISBN{979-8-4007-0327-0/23/12}
\acmConference[ESEC/FSE '23]{Proceedings of the 31st ACM Joint European Software Engineering Conference and Symposium on the Foundations of Software Engineering}{December 3--9, 2023}{San Francisco, CA, USA}
\acmBooktitle{Proceedings of the 31st ACM Joint European Software Engineering Conference and Symposium on the Foundations of Software Engineering (ESEC/FSE '23), December 3--9, 2023, San Francisco, CA, USA}



\begin{document}

\date{}

\title{\underline{M}ate! \underline{A}re You Rea\underline{l}ly A\underline{ware}? An Explainability-Guided Testing Framework for Robustness of Malware Detectors}

\author{Ruoxi Sun}
\affiliation{\institution{CSIRO's Data61}\country{Australia}}

\author{Minhui Xue}
\affiliation{\institution{CSIRO's Data61}\country{}}
\affiliation{\institution{Cybersecurity CRC}\country{Australia}}

\author{Gareth Tyson}
\affiliation{\institution{Hong Kong University of Science and Technology (GZ)}\country{China}}

\author{Tian Dong}
\affiliation{\institution{Shanghai Jiao Tong University}\country{China}}

\author{Shaofeng Li}
\affiliation{\institution{Peng Cheng Laboratory}\country{China}}

\author{Shuo Wang}
\affiliation{\institution{CSIRO's Data61}\country{}}
\affiliation{\institution{Cybersecurity CRC}\country{Australia}}

\author{Haojin Zhu}
\affiliation{\institution{Shanghai Jiao Tong University}\country{China}}

\author{Seyit Camtepe}
\affiliation{\institution{CSIRO's Data61}\country{}}
\affiliation{\institution{Cybersecurity CRC}\country{Australia}}

\author{Surya Nepal}
\affiliation{\institution{CSIRO's Data61}\country{}}
\affiliation{\institution{Cybersecurity CRC}\country{Australia}}

\renewcommand{\shortauthors}{R. Sun, M. Xue, G. Tyson, T. Dong, S. Li, S. Wang, H. Zhu, S. Camtepe, and S. Nepal}

\begin{abstract}
Numerous open-source and commercial malware detectors are available. 
However, their efficacy is threatened by new adversarial attacks, whereby malware attempts to evade detection, \eg by performing feature-space manipulation. 
In this work, we propose an explainability-guided and model-agnostic testing framework for robustness of malware detectors when confronted with adversarial attacks.
The framework introduces the concept of \textit{Accrued Malicious Magnitude (AMM)} to identify which malware features could be manipulated to maximize the likelihood of evading detection.
We then use this framework to test several state-of-the-art malware detectors' ability to detect manipulated malware. 
We find that \one commercial antivirus engines are vulnerable to AMM-guided test cases; \two the ability of a manipulated malware generated using one detector to evade detection by another detector (\ie transferability) depends on the overlap of features with large AMM values between the different detectors; and \three AMM values effectively measure the fragility of features (\ie capability of feature-space manipulation to flip the prediction results) and explain the robustness of malware detectors facing evasion attacks. Our findings shed light on the limitations of current malware detectors, as well as how they can be improved.  
\end{abstract}

\maketitle

\section{Introduction}

The anti-malware market is at the forefront of cybersecurity innovation, constantly driving anti-malware vendors to update their solutions to protect users against a wide range of malicious software variants. The global antivirus software market is expected to reach more than 4 billion USD in 2025~\cite{antivirusMarket}. 
Despite this, recent research has shown that the total number of malware infections has continued to rise~\cite{cyberStatistics}, hitting a new high during the COVID-19 pandemic~\cite{dan,sun2021empirical}.  
There are also numerous high-profile cases of zero-day attacks are being used in offensive campaigns. For example, just a few hours before the Russian-Ukraine conflict, Microsoft's Threat Intelligence Center identified a never-before-seen malware, ``FoxBlade'', that targeted Ukraine's government ministries and financial institutions~\cite{newYorkTimesNews}.

This trend suggests that traditional signature-based and behavior-based methods cannot keep up with the rampant growth of novel malware. Hence, commercial antivirus companies have started using machine learning~\cite{aiAndMl,microsoft} to enable detection without the need for signatures.
However, it has been shown that attackers can evade machine learning-based detectors by manipulating the features that such malware detectors use~\cite{demontis2017yes,demontis2019adversarial,maiorca2015stealth,pierazzi2020intriguing,xu2016automatically}. Because of this, commercial antivirus systems are susceptible to adversarial attacks~\cite{cylance}.
Although there have been numerous works~\cite{goodfellow2014explaining,carlini2017towards,eykholt2018robust,cao2023stylefool,ma2023beatrix} looking at  adversarial attacks in computer vision (where adversaries change specific pixels), adversarial attacks on malware are far less understood. 
We therefore leverage this adversarial approach to build a testing framework to evaluate malware detectors. 
However, we find that it is difficult to fully identify the root causes that impact the decisions of malware detectors.
We therefore reduce this causal discovery problem to identifying the explainable factors that are measurable to impact the robustness of malware detectors.

To implement the testing framework for malware detectors, it is necessary to build techniques that can
\one~generate adversarial malware variants; and \two~measure the explainable features that drive the malware detector's ultimate decision (benign or malicious). 
One existing approach for generating adversarial malware test samples is \emph{obfuscation}.
This focuses on changing the semantic meanings of code snippets in the problem-space (\ie source code), and further obfuscating the malicious signatures or patterns, including hiding the control flow, inserting dummy code, and manipulating variable names~\cite{obfuscapk, vmproject, virbox, jung:avpass-bh, kiteshield, relocbonus}, thereby fooling rule-based malware detectors. 
However, defenses against obfuscation are well-researched. For example, recent research~\cite{chen2020training,avllazagaj2021malware,barbero2022transcending} looks into malware behavior distillation or program behavioral variability analysis towards training robust malware classifiers. Furthermore, Pierazzi~\etal~\cite{pierazzi2020intriguing} argue that the use of mass obfuscation may be counterproductive, rendering antivirus companies to be on the alert~\cite{ugarte2015sok,vigna2018malware}.

In this research, we aim to test the robustness of malware detectors against a new type of adversarial generation: \emph{feature-space manipulation}.
This involves manipulating the malware to trigger changes in the feature space used by detectors. 
Such threats are becoming more prominent because machine learning-based detectors have reduced the efficacy of problem-space attacks.
Feature-space manipulation aims to introduce the intended changes to the feature space (\ie extracted feature vectors) by precisely modifying the problem-space (\ie code).
These manipulating actions could be, for example, adding a redundant section (\eg adding a new code section without linking its address in the section table) or injecting dead code that is unreachable (\eg adding a file I/O request under an always-false condition, so that the dummy code will never be executed). 
Similar techniques have been implemented by Demetrio \etal~\cite{demetrio2021functionality} in a black-box optimization of adversarial Windows malware. 
Although they still focus on problem-space, it is possible to apply such techniques in a feature-space manipulation. 
Due to its urgency, we argue that building a comprehensive testing framework that can automate the identification of limitations in malware detectors is vital.

Several challenges need to be solved during the establishment of a testing framework for robustness of malware detectors.
The \textit{first challenge} is how we can mimic a random attacker in the real world, who may have limited capacity and knowledge.
This means that we are limited to off-the-shelf and easy-to-obtain techniques in the testing.
The \textit{second challenge} is, since most commercial malware detectors are not open-source, this must be done in a detector-agnostic manner.
The \textit{third challenge} is how to interpretively understand the testing results (\eg why test cases work across different detectors).
With the above challenges in mind, to ensure the reproducibility and coverage of our framework, we have several criteria on the development of our testing framework: \one \textit{Easy-to-obtain}, we only utilize open-source and off-the-shelf tools or techniques; 
\two \textit{Model-agnostic}, we decouple the testing strategy from the specifics of the detector; and
\three \textit{Explainable}, an explainable approach is proposed, which will help us to \emph{explain} the root cause of test cases' transferability and identify potential weaknesses in malware detectors.

In this paper, we first propose \textit{Accrued Malicious Magnitude (AMM)} to guide feature space manipulation, finding the correct action(s) on the problem-space that will influence the feature values but without changing run-time functionality.
We then evaluate the robustness of state-of-the-art malware detectors with generated test cases.
To establish such a strategy, we generate test cases by perturbing the feature space and converting the manipulations back into the problem space. 
To achieve our goal, the problem is split into two sub-problems: \one generating test cases; and \two testing malware detectors. 
Our research helps security researchers as well as vendors who develop anti-malware solutions to better understand the robustness of malware detectors and provide insights into how to improve malware defense strategies in an interpretive manner.
The main contributions of this paper are four-fold: 

\begin{itemize}[leftmargin=*]
\item We propose an explainability-guided and model-agnostic testing framework for malware detectors (\S\ref{sec_methodology}). Our framework generates test cases while preserving the malicious functions of the malware. 
We introduce the concept of \textit{Accrued Malicious Magnitude (AMM)} to guide the feature selection approach for feature-space manipulation. 
We further project the manipulated features back into the problem space with a binary builder.

\item We use AMM to test the robustness of state-of-the-art malware detectors (\S\ref{sec_setup}). We show that commercial antivirus engines are vulnerable to AMM-based test cases (\S\ref{sec_eval_perf}). 
Experimental results indicate that feature-space manipulated test cases have significant evasion capability, which decreases the detection rates of 8 state-of-the-art Android malware detectors by \detectionRateDecreaseAvg on average, and bypasses an average of \bypassVTEngines (\bypassVTEnginesPerc) antivirus engines in VirusTotal~\cite{virustotal}. We also highlight the generalizability of our AMM approach by applying it to WinPE malware detectors~(\S\ref{sec_generalizability_analysis}). 
    
\item We explain how manipulations trained on one detector can work on another detector (\ie transferability) through our explaina\-bility-guided approach (\S\ref{sec_transfer}), indicating that this transferability relies on the overlap features that have large AMM values between different machine learning models. 
    
\item We further investigate an approach to improve machine learning-based detectors through excluding \emph{high sensitive but less important} features during training (\S\ref{sec_revisiting_amm_with_improved_detectors}). Results show that AMM values can effectively measure the capability of features of flipping classification results. We suggest that machine learning-based anti-virus products should consider using the AMM values to improve their robustness.  
\end{itemize}

To the best of our knowledge, this is the \emph{first} paper to systematically test the robustness of malware detectors in a way that combines feature-space manipulations with semantic explainability. 

\section{Preliminaries}

\new{In this section, we introduce a motivating example and related research.}

\subsection{Motivating Example}
\begin{figure}[t]
\centering
\includegraphics[width=0.95\linewidth]{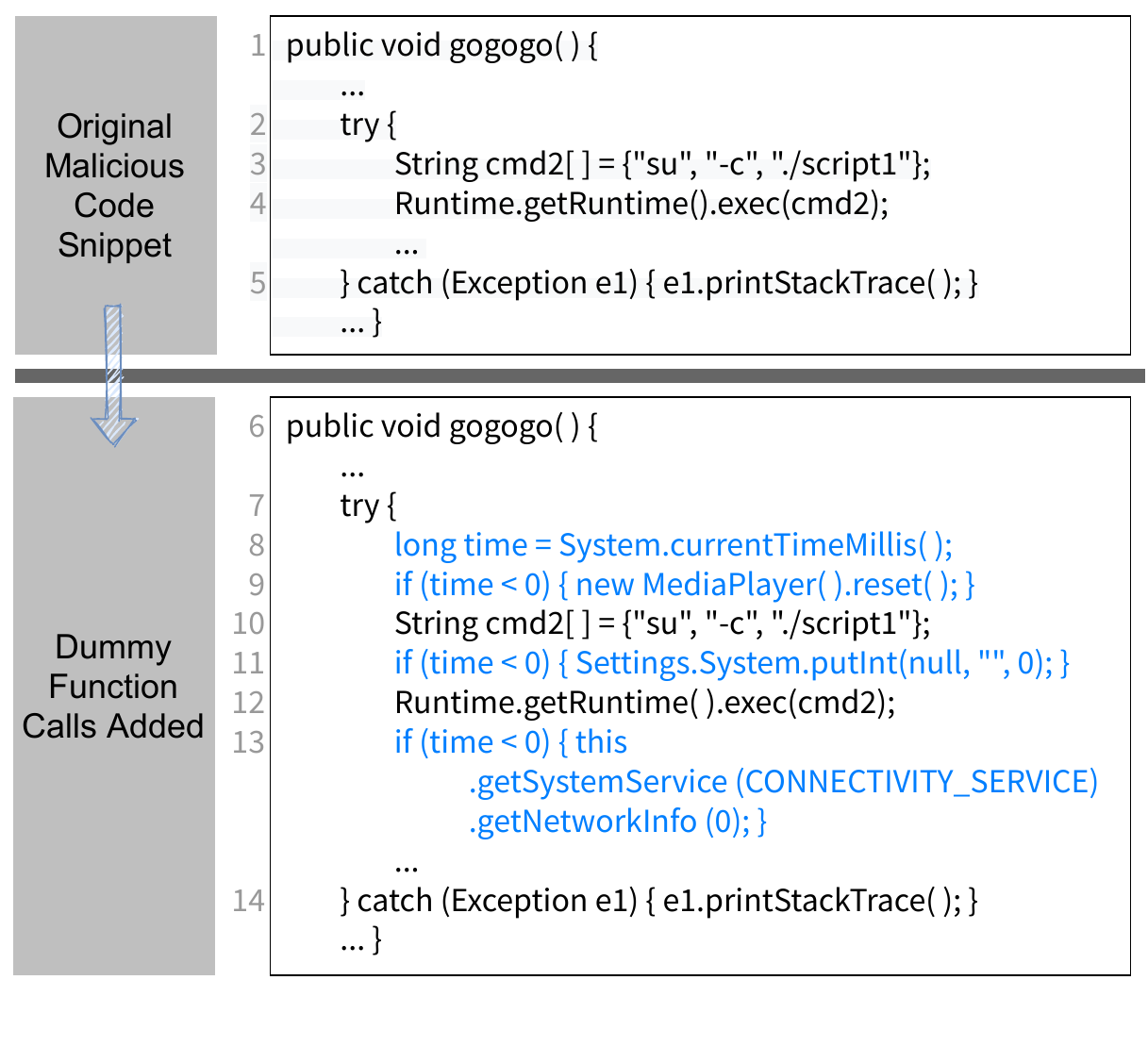}
\vspace{-2mm}
\caption{An intuitive example of injecting dummy `benign' function calls into malicious malware, which bypasses 6 antivirus engines in VirusTotal and evades Drebin.}
\label{fig_motivating}
\end{figure}

\new{To motivate the need for our testing framework, we analyze the source code from an example Android malware, which is tagged as malicious by 39/60 detectors from VirusTotal (VT)~\cite{virustotal}. 
From the source, we find a snippet of malicious code shown in the top part of Figure~\ref{fig_motivating}. As shown in lines 3 and 4, the malware executes a native scripts via root permissions by \texttt{su -c ./script1} command. }

\new{Static features are commonly used in malware detectors where the results is determined through pattern recognition, weighted algorithms, or signature matching.
Thus, it is possible to mislead detectors by introducing more `benign-oriented' elements into malware.
Hence, we insert several `benign' function calls with always-\texttt{false} condition closure (\eg \texttt{time<0}) to ensure they are unreachable during run-time, preserving the original (malicious) functionality. 
After rebuilding the source code, the modified binary is identified by 33 scanners -- 6 fewer than originally, and it bypasses the machine-learning detector provided by Drebin~\cite{arp2014drebin}. The remainder of this paper develops an explainability-guided testing framework and problem-space rebuilding tool that can automate this intuitive idea for the testing of detector robustness.}

\subsection{Related Work}

\noindent \textbf{Malware detectors.~} 
Many modern antivirus engines utilize rule-based analysis, such as signature matching, static unpacking, heuristics matching, and emulation techniques~\cite{malgene, antiviruswork}. 
However, rule-based antivirus engines rely heavily on expert knowledge. With the advantage of feature extraction derived from machine learning techniques, there has been a flurry of work that integrates machine learning models into malware detectors~\cite{arp2014drebin, malgene, kim2018multimodal, DeepRefiner,anderson2018ember, ICCDetector,MaMaDroid,revealDroid}. 
We focus our evaluation on detectors that use static features due to their prevalence in providing pre-execution detection and prevention for many commercial endpoint protection solutions, such as Kaspersky~\cite{kaspersky}, Avast~\cite{avast}, and ESET~\cite{nod32}.

\noindent \textbf{Evaluation of malware detectors.~}
A few studies~\cite{maiorca2015stealth,pomilia2016study} have explored the effect of obfuscations on anti-malware products, utilizing off-the-shelf tools. 
Hammad \etal~\cite{hammad2018large} conducted a large-scale empirical study that evaluates the effectiveness of the top anti-malware products, including 7 open-source, academic, and commercial obfuscation tools. 
Several studies~\cite{shahpasand2019adversarial,chen2019can,suciu2018does} have evaluated machine learning-based malware classifier models with the adversarial samples generated by generative adversarial networks (GANs) or automated poisoning attacks. 
Recent research~\cite{barbero2022transcending,li2021leverage} proposed methods to cope with concept drift or dataset shift, which may lead to performance degradation of malware detectors.
Compared to our research, the scope of these studies only covers either the rule-based products or the machine learning-based models in isolation (rather than both). 

\noindent \textbf{Adversarial samples against malware detectors.~}
The goal of the adversarial attacks is to generate a small perturbation for a given malware sample that results in it being misclassified. This type of attack has been extensively explored in computer vision, and previous research efforts have also investigated the applicability of such techniques to malware classification. 
Xu~\etal~\cite{xu2016automatically} proposed a genetic programming-based approach to perform a directed search for evasive variants for PDF malware. Demetrio~\etal~\cite{demetrio2021efficient} demonstrated that genetic programming based adversarial attacks are applicable to portable executable (PE) malware classifiers. Two recent works~\cite{anderson2018learning, song2021mabmalware} also apply deep reinforcement learning to generate adversarial samples for PE malware to bypass machine learning models. Compared to our research, these studies focus on proposing adversarial attacks rather than testing the robustness of malware detectors and further explaining it. 

\begin{figure*}[t]
\centering
\includegraphics[width=0.9\linewidth]{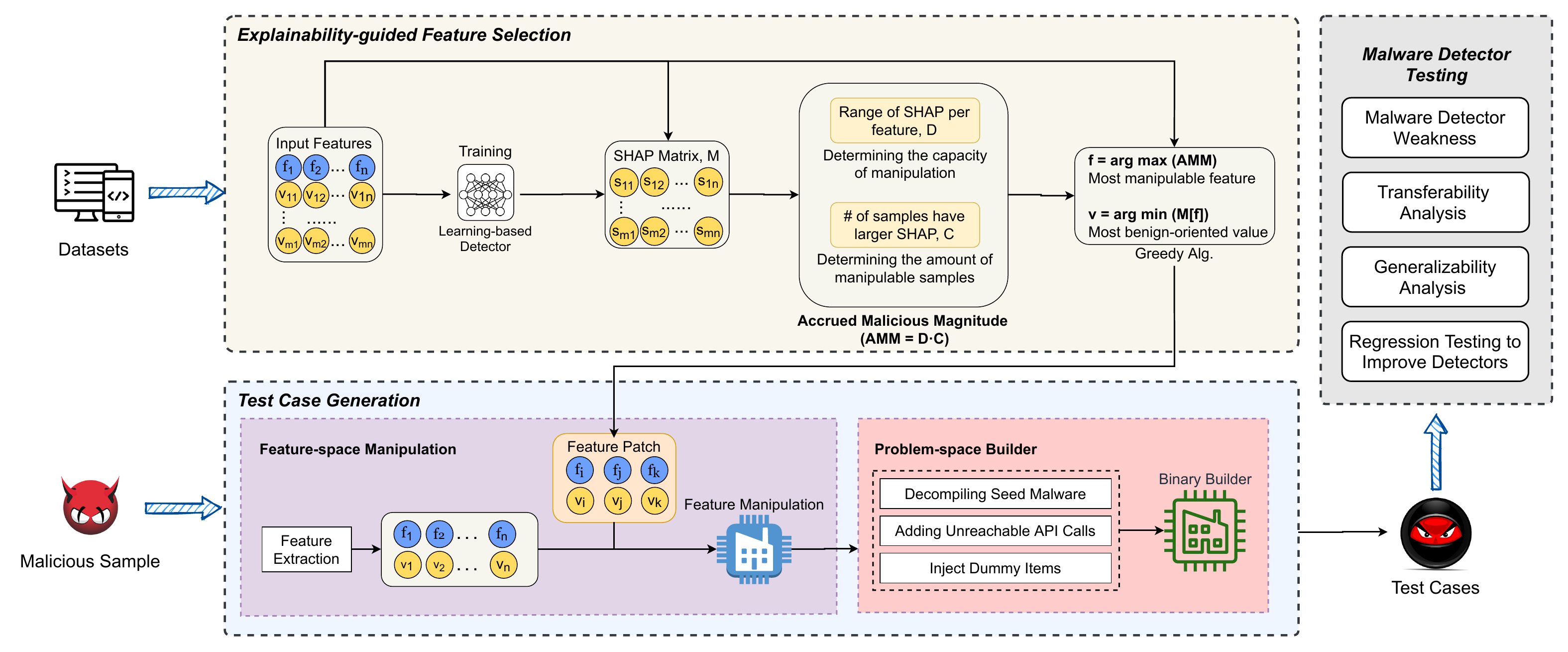}
\caption{The overview of our testing framework.} 
\label{fig_overview}
\vspace{-3mm}
\end{figure*}

\section{Threat Model \& Problem Definition}

In this section, we define the threat model we use in testings, and present our problem definition.

\subsection{Threat Model}\label{subsec_threat_model}
Our testing framework generates adversarial samples to test the robustness of malware detectors.
For this, we must define our assumed threat model. We follow the methodology by Carlini~\etal~\cite{carlini2019evaluating} and describe this threat model with the adversary's goals, capabilities, and knowledge.  

\noindent \textbf{Adversary's goal.~}
The adversary's goal is to manipulate malware samples to evade the detection of malware detectors, including both white-box and black-box detectors. 
In the testing, we only use binary detectors which determine if the software under test is benign or malicious. 
Thus, the goal of attackers is to cause the malicious samples to be misclassified as benign.

\noindent \textbf{Adversary's capability and knowledge.~}
In this work, we assume that an attacker has full knowledge of \emph{one} machine learning model that has been trained for malware detection, including its architecture and training dataset.
This is reasonable as many machine learning models, such as LGBM~\cite{ke2017lightgbm}, are open-source.
Such a white-box model will be used as the source of adversarial sample generation.
\new{With this white-box model, the attacker is capable of ranking the contribution of features and manipulating a malicious sample accordingly to influence its representation in the feature space.}
For other black-box detectors under-test, including machine learning classifiers and antivirus engines, the adversary has no knowledge about the detectors' training dataset, inner structure, or detection mechanism.
For instance, the adversary cannot inject poisoned data into the training dataset or manipulate any code of detectors. 
However, they will still have some basic knowledge about machine learning-based detectors, including access to open-source datasets which could be part of the actual training dataset, generic or popular feature extraction methods~\cite{alinabackdoor}, and off-the-shelf machine-learning detectors.

\subsection{Problem Definition}

Our goal is to test the robustness of malware detectors using the generated test cases.
Consider a malware detector mapping a piece of software ${x \in X}$ to a classification label ${l \in \{0, 1\}}$ (where 0 represents benign and 1 represents malicious). The adversary is trying to mislead this prediction.
Our key test metric is the detection rate on test cases:
\begin{equation}
\begin{split}
    d(x_m) = 1, ~x_t = \Gen(x_m), ~d(x_t) = 0, \\
    \performance(d) = \frac{1}{n}\sum\limits_{i = 1}^n d(x_t^i)
\end{split}
\label{equ:classifier}
\end{equation}
\noindent where $d$ is the malware detector which could be either a  machine learning detector (a feature extraction method plus a trained model) or an antivirus engine. $x_m$ is the original malware sample, and $\Gen$ is the generator of test case, $x_t$, while keeping its malware functionality the same as $x_m$. $\performance(d)$ is defined as a detector's detection rate on $n$ test cases, indicating the robustness of $d$. 

\section{Testing Framework}\label{sec_methodology}

Our testing framework consists of three key components (see Figure~\ref{fig_overview}): 
\one explainability-guided feature selection, to select the features for manipulation; 
\two a test case generator that relies on the previously selected features;
and 
\three malware detector testing to identify which detectors are robust against the adversarial malware samples. Before diving into the testing framework, we would like to introduce preliminary knowledge about SHAP~\cite{lundberg2017unified}, the model explanation technique we used in our testing.

\subsection{A Primer on SHAP}\label{apdx_shap}
Research into explainable machine learning has proposed multiple systems to interpret the predictions of complex models. 
We rely on SHAP~\cite{lundberg2017unified} (based on the coalitional game theory concept of Shapley values) as prerequisite knowledge to bootstrap the testing framework. 
The SHAP framework subsumes several earlier model explanation techniques together, including LIME~\cite{ribeiro2016should} and Integrated Gradients~\cite{sundararajan2017axiomatic}.
SHAP has the objective of explaining the final value of a prediction by attributing a value to each feature based on its contribution to the final result. 
To accomplish this task, the SHAP  frameworks train a surrogate linear explanation model $g$ of the form:
\begin{equation}
\begin{split}
    f(x) = g(x'),\\
    g(x') = \phi_0 + \sum\limits_{j=1}^M{\phi_jx_j'},
\end{split}
\label{equ:shap}
\end{equation}
\noindent where $f$ is the original model, $x$ is the input sample to be attributed, $x^\prime$ is the coalition vector of $x$.
$\phi_0=\mathbb{E}_X(f(X))$ is the average prediction of the original model on sampled dataset~$X$. The Shapley value $\phi_j \in \mathbb{R}$ is the feature attribution for the $j^{th}$ feature $x_j'$ to the model’s decision. 
Summing the effects of all feature attributions approximates the difference of prediction for $x$ and the average of the original model, aiming to explain any machine learning-based model without internal knowledge.

LIME uses a linear explanation model $g(x')$ to locally approximate the original model, where locality is measured in the simplified binary input space, \ie $x' \in \{ 0, 1 \}^M$. 
To find $\phi$, LIME minimizes the following objective function:
\begin{equation}
\begin{split}
    \xi = \underset{g \in G}{\argmin}\ L(f, g, \pi_{x}) + \Omega(g),
\end{split}
\label{equ:shap_lime}
\end{equation}
\noindent where $L$ is the squared loss over a set of samples in the simplified input space weighted by the kernel function $\pi_{x}$, and $\Omega$ penalizes the complexity of $g \in G$ where $G$ is hypothesis space. Therefore, based on the input feature vectors and the output predictions of the model, in this research, we use SHAP to approximate the fragility of each feature, selecting features that can be manipulated to flip the prediction.

\subsection{Step 1: Feature Selection}

In the first step of our methodology, we utilize SHAP to create explainability-guided adversarial test cases.
In this step we use a single detector model to generate SHAP values for the input dataset, calculating how much one feature contributes to an individual prediction.
It is assumed that an adversary would have access to this single model. 
The workflow of the explainability-guided feature selection is illustrated in Algorithm~\ref{alg_adv_greedy}. For a set of seed malware, $X_s$, we generate a corresponding test case set, $X_t$. 

\noindent \textbf{Pre-processing.~} We extract features from the training samples $X$ of a trained machine learning model $m$ (line 2). 
Here we will use a generic feature extraction approach that is adopted from open-sourced detectors to mimic the adversary's capability and knowledge. 
\new{Using Android as an example, Table}~\ref{tab_feature_extraction} \new{summarizes the feature extraction methods of several state-of-the-art for Android malware detection. We adopt features that are representative and easy to extract. 
It is possible to generate stronger evasive adversarial test cases with more features involved. However, our goal is to evaluate malware detectors with an easy-to-obtain approach, which should provide a lower bound of robustness.
Note, we use a similar strategy to determine the feature extraction methods for other operating systems, as described in} \S\ref{sec_generalizability_analysis}.

\begin{table}[t]
\centering
\caption{Feature extraction methods used in different Android malware detectors and in our measurement.}\label{tab_feature_extraction}
\resizebox{\linewidth}{!}{
\begin{threeparttable}
\begin{tabular}{@{}lccccc@{}}
\toprule
\textbf{Features} & 
\textbf{\begin{tabular}[c]{@{}c@{}}Drebin\\ ~\cite{arp2014drebin}\end{tabular}} & 
\textbf{\begin{tabular}[c]{@{}c@{}}MaMaDroid\\ ~\cite{MaMaDroid}\end{tabular}} & 
\textbf{\begin{tabular}[c]{@{}c@{}}RevealDroid\\ ~\cite{revealDroid}\end{tabular}} & 
\textbf{\begin{tabular}[c]{@{}c@{}}DroidSpan\\ ~\cite{droidSpan}\end{tabular}} & 
\textbf{\begin{tabular}[c]{@{}c@{}}Our \\ Measurement\end{tabular}} \\ \midrule
\textbf{Permissions} & \pie{360} &  &  &  & \pie{360} \\
\textbf{Hardware Components} & \pie{360} &  &  &  &  \\
\textbf{App Components} & \pie{360} &  &  &  &  \\
\textbf{API Calls} & \pie{360} & \pie{180} & \pie{360} & \pie{360} & \pie{360} \\
\begin{tabular}[c]{@{}l@{}}\textbf{Strings} \\ (\eg Network Addresses)\end{tabular} & \pie{360} &  &  &  & \pie{360} \\
\textbf{Call Graphs} &  & \pie{360} &  & \pie{360} &  \\
\textbf{Native Call} &  &  & \pie{360} &  &  \\ \bottomrule
\end{tabular}
\begin{tablenotes}
\item[] \pie{360}: the feature is involved; \pie{180}: the feature is indirectly involved.
\end{tablenotes}
\end{threeparttable}
}
\end{table}

Then the vectorized samples~$X^\prime$ and the model are input to $\shap()$ to calculate the SHAP value matrix $M$ (line 3).
The matrix is then used to select the most evasive features and the most benign-oriented values (\ie the most negative values that exist in the selected features, as 0 represents benign). 

\noindent \textbf{Feature selection.~}
To select the feature that has largest malicious magnitude, we propose the concept of \textit{Accrued Malicious Magnitude (AMM)}.
The AMM is defined as the product of the magnitude of SHAP values in each feature and the number of samples that have malicious-oriented values (\ie values towards the positive side, as 1 represents malicious) in the corresponding feature. 
By calculating AMM values, we select the feature that has the largest modifiable capability and has the most samples to be modified as the test cases, \ie samples that have SHAP values towards the positive (malicious) side, which have the potential to be manipulated to benign. 
Specifically, starting from the $\range(M)$ in line 5, we first calculate the range of SHAP values in each feature and store the results in a one-dimension vector $D$. $D$ indicates the potential magnitude we can modify on each feature, \ie each $d_i \in D$ presents the difference between the maximum SHAP value and the minimum SHAP value of feature $f_i$. Next, for each feature, we count how many samples have a SHAP value larger than the mean SHAP value of that feature (the $\countlarge(M)$ in line 6) and collect the results in a one-dimensional vector $C$. 
Therefore, a larger $c_i \in C$ means that, for feature $f_i$, there are more samples that have a SHAP value towards malicious, such that more samples can be manipulated towards benign. Therefore, we select the most evasive feature according to the AMM values, denoting the dot product of the range of SHAP values~($D$) and the number of SHAP values greater than mean ($C$) (line~7). 

\noindent \textbf{Value selection.~}
Once we have identified the feature $f$ to compromise, the next step is to choose the value for the selected feature to guide the manipulation. 
We select the most benign-oriented value, $v$, in the feature space. This corresponds to the most \textit{\textbf{negative}} value in $M[f]$, the SHAP value of the feature~$f$~(line 9). 

\noindent \textbf{Updating the feature.~} After obtaining $(f, v)$, if the selected feature $f$ is manipulable, we add the pair into a map, $P$, as the \emph{Feature Patch} to be used in the feature-space manipulation (line 11). 
Note that, due to the strong semantic restrictions of the binaries, we cannot simply choose any arbitrary pairs of feature and values for the test manipulation. Instead, we restrict the feature-space manipulation to only features and values that are independent (IID) and can be modified with original functionalities preserved.
For example, consider the feature that counts the size of a binary: When we modify the value of another feature, the former will be modified indirectly.
Therefore, the features and values we select to be manipulated follow two principles employed by the previous literature~\cite{alinabackdoor, grosse2016adversarial, grosse2017adversarial}. 
These principles are:
\one features are manipulable in the original problem space; and
\two selected features have no dependencies or cannot be affected by other features. We described manipulable features for Android and WinPE datasets later in \S\ref{sec_dataset} and \S\ref{sec_generalizability_analysis}.

\noindent \textbf{Greedy strategy.~}
After obtaining feature-value pairs, we conduct a greedy strategy, removing samples that have the same value, $v$, for feature $f$ from the dataset (lines 13 to 17). 
We do this to make sure that the same feature-value pair will not be selected again. The procedure repeats until we find $N$ feature-value pairs. These $N$ pairs are then used as feature patch in the next stage to generate test cases.

\begin{algorithm}[t]
\footnotesize
\SetAlgoLined
\KwIn{Machine learning model $m$, dataset $X$, and the number of features to be selected $N$.}
\KwOut{Feature patch $P$.}
$P = \map(Feature, Value)$\;
$X' \gets \vectorize(X)$\;
$M \gets \shap(X',m)$\;

\While{$size(P) < N$}{
$D \gets \range(M)$\;
$C \gets \countlarge(M)$\;
$AMM \gets D \cdot C$\;
$f \gets \argmax(AMM)$\;
$v \gets \argmin(M[f])$\;
\If{$\ismanipulable(f)$}{
$P \gets P \cup (f,v)$\;
}

\For{$each\ x' \in X'$}{
\If{$x'[f] \neq v$}{
$idx \gets \getindex(X',x')$\;
$M \gets M \setminus M[idx]$\;
$X' \gets X' \setminus x'$\;
}
}
}
\Return $P$;

\caption{ \bf AMM-based Feature-Space Selection}\label{alg_adv_greedy}
\end{algorithm}

\subsection{Step 2: Test Case Generator}\label{sec_adv_smp_gnr}

In the next step, the test case generator conducts feature manipulation according to the feature patch obtained from prior steps.  

\noindent\textbf{Feature-space manipulation.~}
Equation~\ref{equ:shap2} summarizes the feature-space manipulation: 
\begin{equation}
\begin{split}
    x_m' = \vectorize(x_m), \\
    x_t' = \manipulate(x_m', P), \\
    x_t = \Gen(x_m, x_t'), 
\end{split}
\label{equ:shap2}
\end{equation}
\noindent where $x_m^\prime$ is the result of applying feature extraction on a malicious sample~$x_m$ using $\vectorize()$. 
$\manipulate()$ manipulates the sample in feature-space guided by the selected feature and value pairs, $P$. Note that, the sample generator $\Gen()$ will take the manipulated feature-space sample $x_t'$ and the original seed sample $x_m$ as input, and implement the changes in feature-space back to problem-space to generate the test case, while keeping its malware functionality (as detailed in Binary builder below).

\noindent \textbf{Binary builder.~} 
To ensure that no loss of functionality is inadvertently introduced as a side effect of feature manipulation, we only apply these changes to unreachable areas of binaries, so that these changes will never be executed during run-time. Therefore, we guarantee that test cases are executable and can be applied in the testing of malware detectors in the wild. Then, we apply these changes on seed binaries with the help of open-source binary builders. 

For simplicity, we present our tooling for Android malware, yet we emphasize that our framework works with other operating systems (see \S\ref{sec_generalizability_analysis}).
We adopt a similar feature extraction method of Drebin on the Android APK. Since features are a vector of boolean values representing the existence of a feature, the feature value could only be modified from 0 (absence) to 1 (presence) to preserve original functionalities. 
We first leverage Apktool~\cite{apktool} to decompile an APK file into Smali~\cite{smali} code, a structured assembly language. 
API calls and network URLs are transformed to smali instruction code, which is wrapped by an unreachable disclosure, \eg an always-false condition closure such as \texttt{if(time < 0)}. The Smali code is then inserted into the Smali file of the main activity. Features representing Android manifest components are inserted into AndroidManifest.xml file directly. Finally, we utilize Apktool to assemble all decompiled and manipulated files into an adversarial APK sample. 
If a feature manipulation cannot be implemented in this way, we skip it and continue with the next feature in the selected patch. 

\subsection{Step 3: Malware Detector Testing}\label{sec:malware_evaluation}

After the test cases are generated, our framework programmatically executes a series of tests on the malware detectors. This involves
\one~calculating the per-detector robustness;
and
\two~testing the transferability of attacks across detectors. 
We conclude this subsection by proposing techniques that can improve detector performance.

\noindent \textbf{Testing detector robustness.~}
The methodology to test detector robustness is straightforward.
We first input the seed malware into each detector and collect the detection rate on the unaltered malware dataset.
Next, the test cases generated from the white-box model will be input to the detectors (including the white-box model itself). 
We then compare the difference between the detection rate of the seed \vs test cases.
This allows us to test which detectors are vulnerable to test cases, \ie whether the manipulation on AMM features changes the detector results. 

\noindent \textbf{Transferability analysis.~}
Considering that machine learning-based detectors may use similar feature extraction methods, it is possible that multiple detectors are susceptible to the same feature manipulations. Specifically, we posit the test cases generated from one machine learning model are likely to be effective on other models that are trained on the same data distribution, due to the similarity of decision boundaries.
Therefore, the test framework also calculates the ability to transfer an evasion trained on one detector to another, \ie transferability.
This is important as, the more powerful the transferability a malware test case has, the more effective it is against other detectors. To evaluate the transferability of test cases generated by the AMM-based approach, we generate cases from white-box models and apply them to black-box models. If transferability exists, the test framework calculates the feature-space overlaps among models to explore the root cause of transferability. This can be used by developers to understand the weaknesses in their feature engineering.

\noindent 
\textbf{Regression testing for detector improvement.~}
Inspired by recent research~\cite{ilyas2019adversarial}, a detection model can be improved by removing important features from the training phase, where the important features refer to the ones that are highly contributing to the model prediction accuracy. 
Shapley Additive Global importancE (SAGE)~\cite{covert2020understanding} is a framework that measures how much a feature contributes to the prediction accuracy of a model. 
We apply the improvement on the detection models with SAGE and test their robustness against adversarial samples. Specifically, we first calculate SAGE values of each feature. Since the most important features are the ones with largest SAGE values, we sort the features by SAGE values in a descending order and select the top features. Then we remove the top features from samples and generate a new training set. Finally, an improved model $m_i$ is trained with the new training set. 
To establish a thorough comparison, we train another model $m_a$ that excludes top AMM-based features to compare and explain the improvement of malware detectors.

\section{Testing Framework Setup}\label{sec_setup}

In this section, we describe the setup of our tests, including the detectors under test, the datasets, and how we trained the models. 

\subsection{Detectors under Test}
To showcase our testing framework, we experiment with a number of malware detectors.
For Andoird malware detectors, we test the robustness of 8 state-of-the-art machine learning-based detectors (a combination of 2 feature extraction methods and 4 machine learning models); 
for WinPE detectors, we test 4 detectors (1 feature extraction method accompany 4 models). We also test 60 antivirus engines available using VirusTotal~\cite{virustotal}. The detectors under test are listed in Table~\ref{tab_detectors}. Specifically, we involve Drebin~\cite{arp2014drebin}, MaMaDroid~\cite{MaMaDroid}, and Ember~\cite{anderson2018ember} in our testing, because 
\one they proposed unique features extraction methods; 
\two high accuracy rates are reported on large datasets; 
and 
\three their source code and datasets have been made publicly available.

To follow the conventions of prior studies~\cite{alinabackdoor,arp2014drebin,MaMaDroid,revealDroid,droidSpan}, we select 4 off-the-shelf machine learning-based malware models that are commonly used in malware detectors. 
In our threat model, the adversary has full knowledge of one machine learning-based model.
Without loss of generality, we set LGBM as this \textit{white-box} model. 
Note, any machine learning model could serve this role as our approach is model-agnostic. For the antivirus engines, we use VirusTotal, an online service that provides over 70 antivirus scanners to detect malicious files and URLs. We find that 60 scanners are always available while the others are not stable. Therefore, we include these 60 scanners in our evaluation. All antivirus engines fall into the black-box category as the attacker has no specific knowledge about them. 


\begin{table}[t]
\caption{Detectors under test.}\label{tab_detectors}
\resizebox{0.9\linewidth}{!}{
\begin{tabular}{@{}p{1.5cm}lp{7cm}@{}}
\toprule
\textbf{Type} & \textbf{Name} & \textbf{Description} \\ \midrule
\multirow{9}{1.5cm}{Feature Extraction Methods}
 & \multirow{3}{1.5cm}{Drebin \cite{arp2014drebin}} & A lightweight method for Android malware detection, extracting 8 sets of features from an application’s code and manifest.  \\ \cmidrule{2-3}
 & \multirow{3}{1.5cm}{MaMa\-Droid \cite{MaMaDroid}} & A static-analysis-based system that abstracts app’s API calls and builds a model from the call graph of an app as Markov chains. \\ \cmidrule{2-3}
 & \multirow{3}{1.5cm}{Ember \cite{anderson2018ember}} & A static WinPE malware classifier extracting eight groups of raw features that include both parsed features and format-agnostic histograms and counts of strings. \\ \midrule
\multirow{9}{1.5cm}{Machine Learning Models}
 &\multirow{2}{1.5cm}{LGBM \cite{ke2017lightgbm}} & LightGBM, an open-sourced gradient boosting framework, based on the decision tree algorithm. \\ \cmidrule{2-3}
 & \multirow{2}{1.5cm}{SVM} & Support Vector Machine, a  supervised learning method based on statistical learning frameworks. \\ \cmidrule{2-3}
 & \multirow{2}{1.5cm}{RF} & Random Forests, an ensemble learning method that combines decision trees to provide classification. \\ \cmidrule{2-3}
 & \multirow{3}{1.5cm}{DNN} & A feed-forward neural network with  with one input layer and three fully-connected hidden layers (the last one ends with a Softmax function).\\ \midrule
\multirow{4}{1.5cm}{Antivirus Engines}
 & \multirow{4}{1.5cm}{VT~\cite{virustotal}} & VirusTotal, a website that aggregates more than 70 antivirus products and online scan engines, allowing a user to check for viruses that the user's own antivirus software may have missed. \\ \bottomrule
\end{tabular}
}
\end{table}

    
    
    

\subsection{Datasets and Model Training} \label{sec_dataset}
In our experiments, we conduct testings using an Android dataset and a WinPE dataset.

The Android Application Package (APK) is the package file format used by the Android operating system for distribution of mobile apps. We use the well-studied Drebin~\cite{arp2014drebin} dataset, which contains 123,453 benign samples and 5,560 malicious. We also include 8,351 malware samples that have been uploaded on VirusShare~\cite{virusshare}. 
Since the ratio of malicious and benign apps is unbalanced, we randomly select 5,560 benign and 5,560 malicious samples, making up 11,120 samples. Further, we create a random 50:20:30 split of samples for training, validation, and testing, to train a LightGBM model. 3,000 test cases are generated from 3,000 randomly selected malicious samples. The details about the WinPE dataset establishing and model training are provided in \S\ref{sec_generalizability_analysis}.

To train the ML models mentioned above, we employ Androguard~\cite{androguard} to extract raw features from APK samples. Androguard is a python tool to analyze and manipulate Android files. 
It disassembles an APK file and converts its byte code and resource files into a readable and structured format. We further extract features from the manifest file and from the Dalvik Executable (dex) file. These features are then used to train and evaluate the machine learning based detectors. All Android features are manipulable as they are independent to each other.

\begin{figure*}[t]
\centering
\begin{minipage}[t]{0.28\linewidth}
\includegraphics[width=\linewidth]{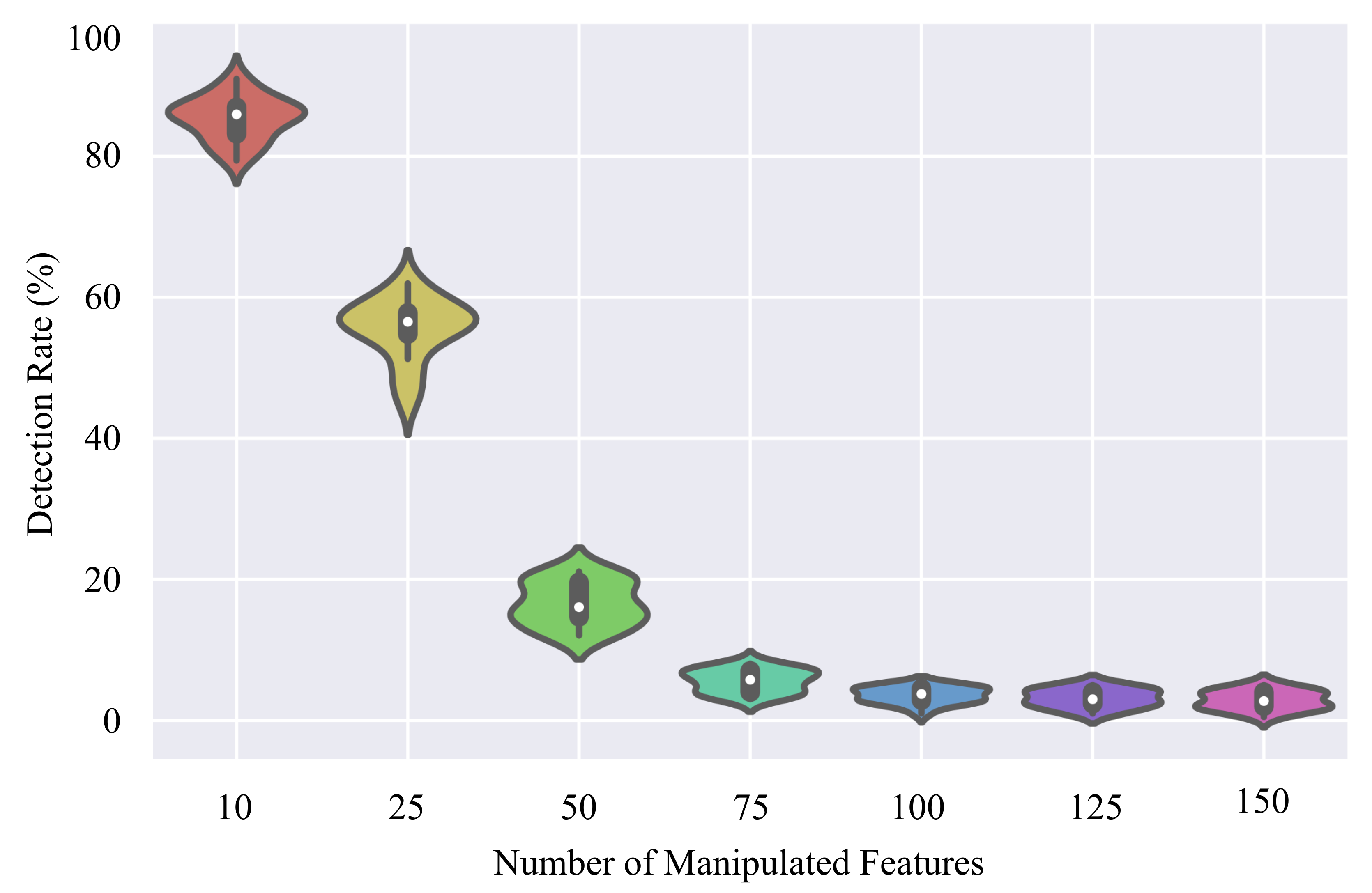}
\caption{Detection rates of manipulating different numbers of features.}
\label{fig:shap_compare}   
\end{minipage}
\begin{minipage}[t]{0.65\linewidth}
\includegraphics[width=\linewidth]{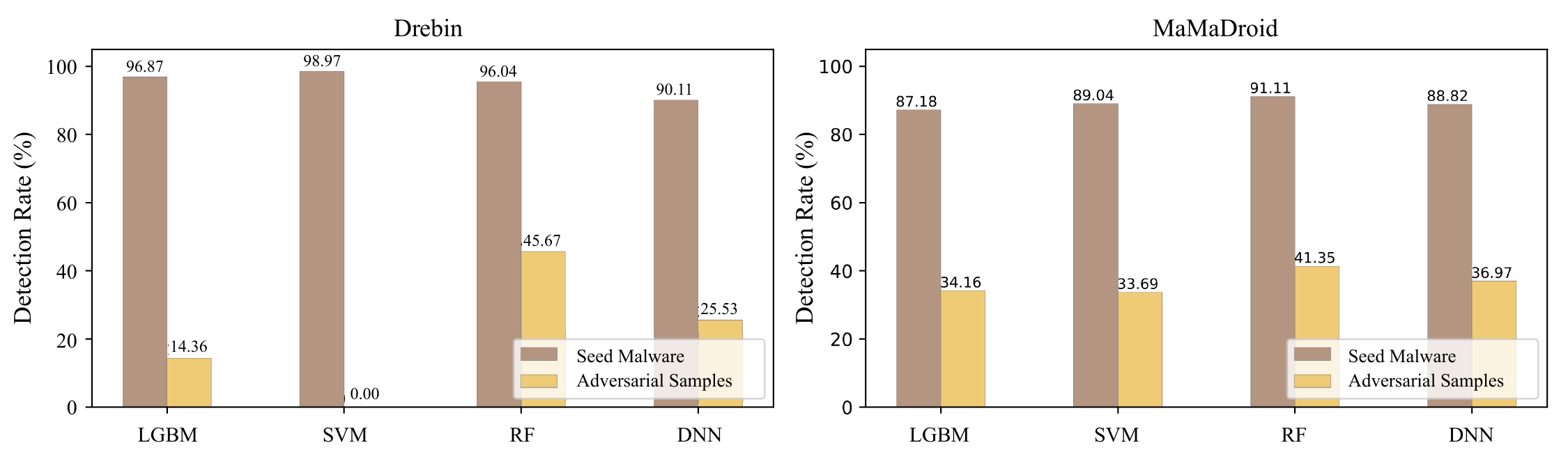}
\caption{Testing results on machine learning-based detectors.}
\label{fig_main_study}
\end{minipage}
\vspace{-2mm}
\end{figure*}

\section{Testing Framework Results}
\label{sec_results}

We next employ our explainability-guided test framework to evaluate the robustness of the detectors.
Specifically, we compare the detection rates between \emph{original} samples and the \emph{test cases} generated by our proposed strategy. To choose the number of features to manipulate, we conduct pilot experiments on Android and WinPE datasets.

\subsection{Pilot Experiment}\label{pilot_experiment}

\new{\noindent \textbf{Number of Features.~}To choose the number of features to manipulate in the explainability-guided test case generation, we conduct pilot experiments on the Android and WinPE datasets. Here we only report the experiment on the Android dataset as they follow the same strategy. 
The pilot experiment compares the detection rate of the LGBM model while manipulating 10, 25, 50, 75, 100, 125, and 150 features on 20 sets of 100 seed samples (2,000 samples in total). As shown in Figure}~\ref{fig:shap_compare}, \new{with more features manipulated, the detection rate decreases. The trend turns at 75 features, with 5.1\% test cases detected. We therefore choose $N=75$ for the APK binaries. }

\begin{figure}[t]
\begin{minipage}[t]{.48\linewidth}
\includegraphics[width=1\linewidth]{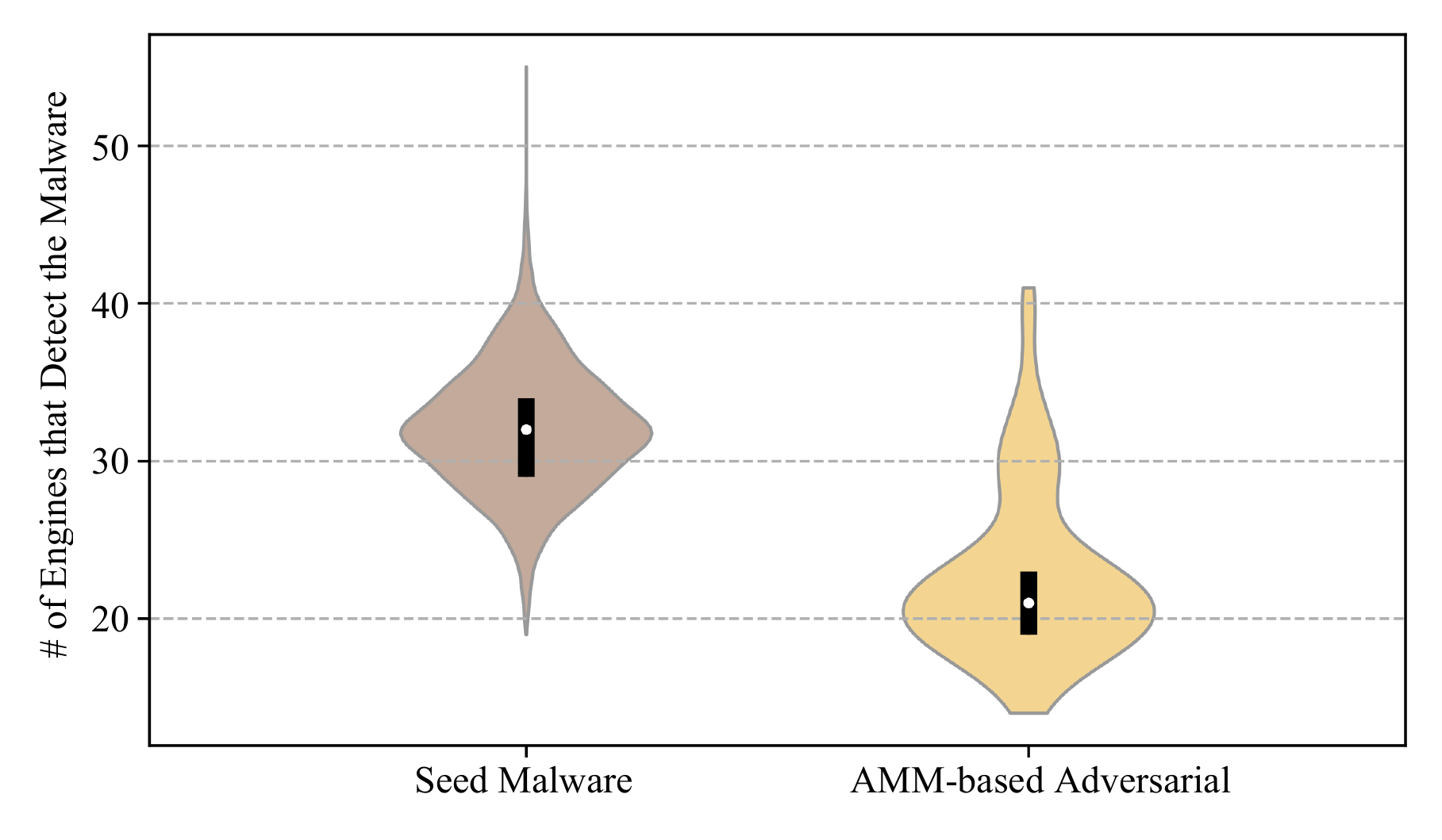}
\captionof{figure}{Number of VirusTotal anti-virus engines that can detect original malware samples and test cases.}
\label{fig_av_results}
\end{minipage}
\hfill
\begin{minipage}[t]{.48\linewidth}
\includegraphics[width=1\linewidth]{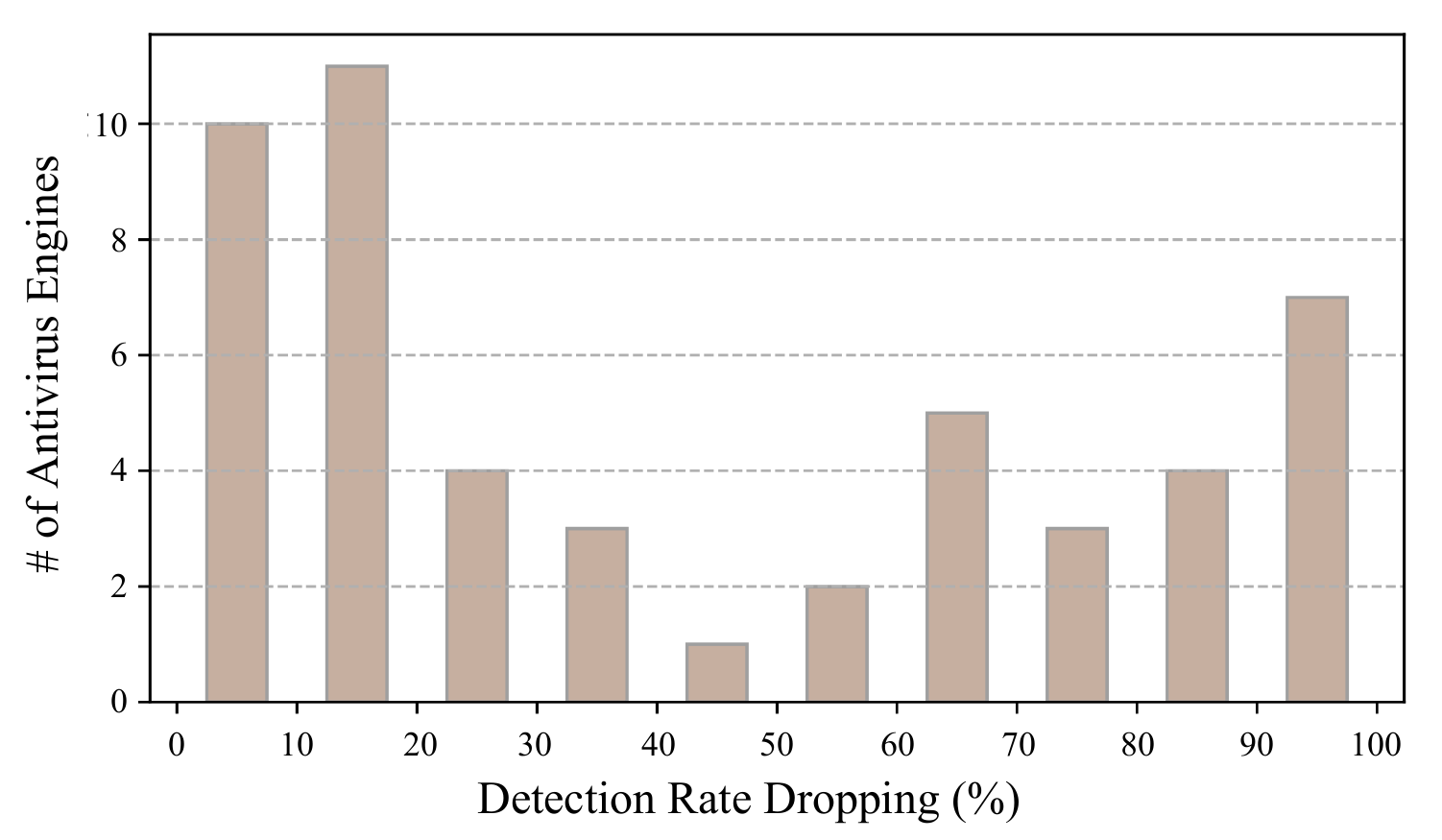}
\captionof{figure}{Histogram of antivirus engines with respect to the detection rate decreasing.}
\label{fig_dropping}
\end{minipage}
\vspace{3mm}
\end{figure}

\subsection{Testing Results on Android Malware Detectors}
\label{sec_eval_perf}

In our Android malware detector testing, we generate APK test cases from the LGBM model, and test on 8 machine learning-based detectors using 2 feature extraction methods and 4 models (1 in white-box and 3 in black-box), and 60 antivirus engines (black-box) from VT. 

\noindent \textbf{Testing machine learning detectors.}
All machine learning-based detectors have reasonable detection rate (above 85\% on average) on the original malware samples, as shown in Figure~\ref{fig_main_study}.
Note, the y-axis refers to the detection rates of samples and each detector is presented on the x-axis.
On the Drebin-based detectors, the test cases averagely reduce the detection rate by \detectionRateDecreaseDrebinPercAvg (ranging from 50.37\% on Drebin-RF to 98.97\% on Drebin-SVM). Noticeably, all test cases bypass the detection of Drebin-SVM. This may be attributable to \one our feature extraction method being close to Drebin; and \two the SVM, being a simple linear model, making its prediction easier to flip. We further analyze the robustness of the SVM model as a case study in \S\ref{sec_case_studies}. While on MaMaDroid-based detectors, considering that MaMaDroid utilizes call graph features, which are indirectly related to our API call features extracted from Android source code, it is expected that fewer test cases can bypass the detection. Specifically, the detection rates of MaMaDroid-based detectors decrease \detectionRateDecreaseMaMaPercAvg on average (ranging from 49.76\% on MaMaDroid-RF to 55.35\% on MaMaDroid-SVM) on test cases. The detection rates decrease down to \detectionRateMaMaPercAvg (avg.) which is even lower than random guess. We conclude that all 8 detectors involved in our Android malware detector testing are vulnerable to the test cases generated by the explainability-guided feature space manipulation.  
Note, in white-box scenarios (\ie the 2 detectors using LGBM model), it is expected that the generated test cases are more effective against white-box models.
To further explore the reason why our explainability-guided approach also works in black-box scenarios, we present further test results on transferability in \S\ref{sec_transfer}.

\noindent \textbf{Testing VirusTotal engines.}
We next turn our attention to the 60 antivirus engines accessible via VirusTotal.
Figure~\ref{fig_av_results} plots the number of virus engines that successfully flag the seed malware and test cases as malicious.
The average detection rate of seed malware is \vtScoreOriPerc, \ie on average \vtScoreOri VT detectors detect a sample as malicious. In contrast, only an average of \vtScoreAdv  detectors flag test cases as malicious.
This lowers the detection rate of VT to \vtScoreAdvPerc on average.   
This result indicates that not all detectors in VT are robust against these test cases.

To illustrate the performance of each antivirus engine and explain the overlapping results between samples, Figure~\ref{fig_dropping} presents a histogram of the detection rate reductions for each antivirus engine.
The x-axis represent the range of detection rates decreasing and the y-axis refers to the numbers of antivirus engines. 
To obtain a meaningful result, 10 engines that have detection rates less than 50\% on seed malware samples are excluded. 
According to Figure~\ref{fig_dropping}, 10 engines are relatively resistant to the test cases, with detection rate reductions of under 10\%. 
This explains the overlapping results between the upper percentile of test cases and the mean of the seed malware samples. 
In contrast, there are 22 engines with detection rates that decrease by over 50\%, indicating that these antivirus engines are vulnerable to the AMM-based test cases. 

\vspace{1mm}\begin{mdframed}[backgroundcolor=black!10,rightline=false,leftline=false,topline=false,bottomline=false,roundcorner=2mm,everyline=true,nobreak=false] 
\textbf{Takeaway 1:}  
\begin{itemize}[leftmargin=*]
\item Machine learning-based detectors are vulnerable to the test cases generated by the AMM-based strategy, which manipulates the most evasive features.
\item Test cases can evade detection by the black-box antivirus engines in VirusTotal. This indicates that the majority of anti-malware vendors could use our testing framework to examine and improve their detectors.
\end{itemize}
\end{mdframed}

\subsection{Transferability Analysis}\label{sec_transfer}

Transferability is the ability for a test case to be effective against multiple learning-based detectors. To study this, taking Drebin as an example, we next generate AMM-based test cases from each of the detectors and input them to different detectors. Here, we seek to understand how a manipulation guided by one detector performs against other detectors.\footnote{As the calculation of SHAP matrix on DNN model requires large amount of computing time using our facility, we conduct transferability analysis on 3 generation models and we believe such experimental setting is sufficient.} 
Specifically, we measure the transferability in two aspects: feature overlaps and detection rates.

\begin{figure}[t]
\centering
\includegraphics[width=0.8\linewidth]{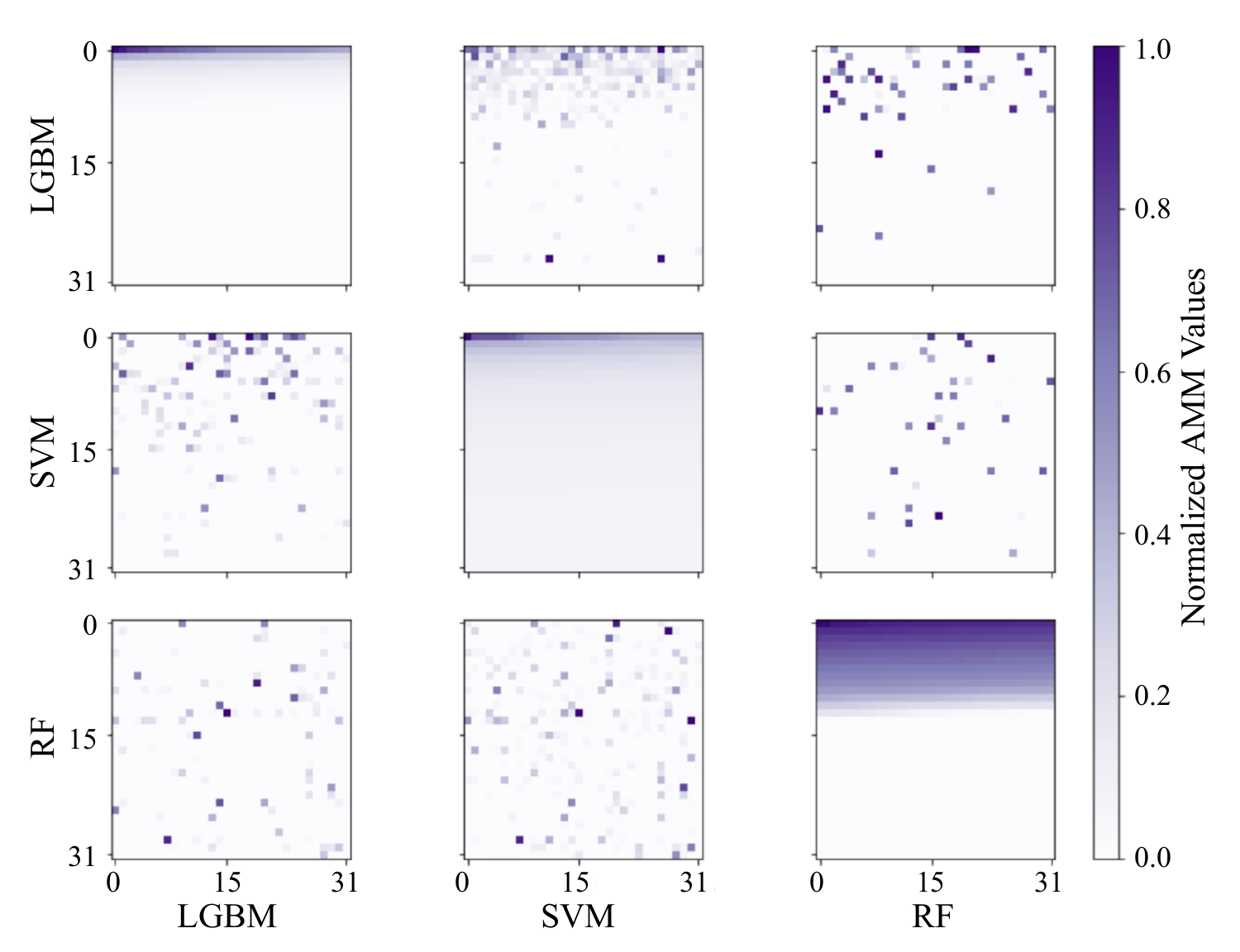}
\caption{AMM values of top 1,024 features that have the highest AMM values in LGBM (row 1), SVM (row 2), and RF (row 3) models, separately. } 
\label{fig_shap_heatmap}
\end{figure}

\begin{figure*}[t]
\begin{minipage}[t]{.3\linewidth}
\includegraphics[width=1\linewidth]{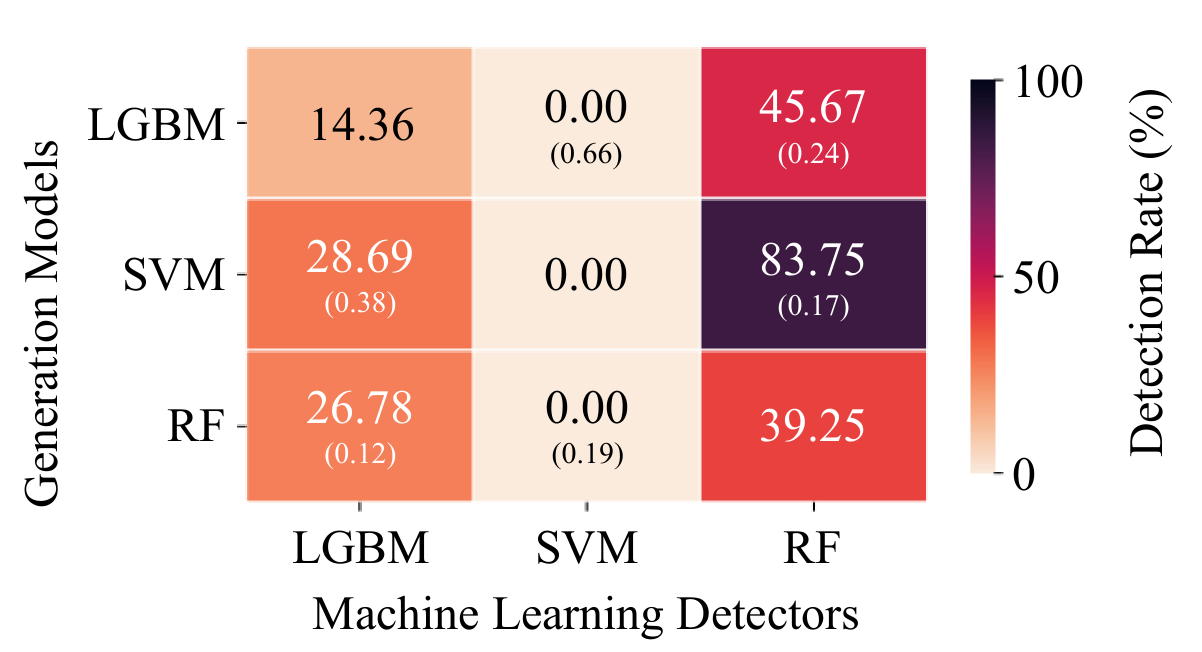}
\captionof{figure}{Detection rates of three Drebin-based detectors on test cases generated from LGBM, SVM, and RF.} 
\label{fig_transferability_heatmap}
\end{minipage}
\hfill
\begin{minipage}[t]{.3\linewidth}
\includegraphics[width=1\linewidth]{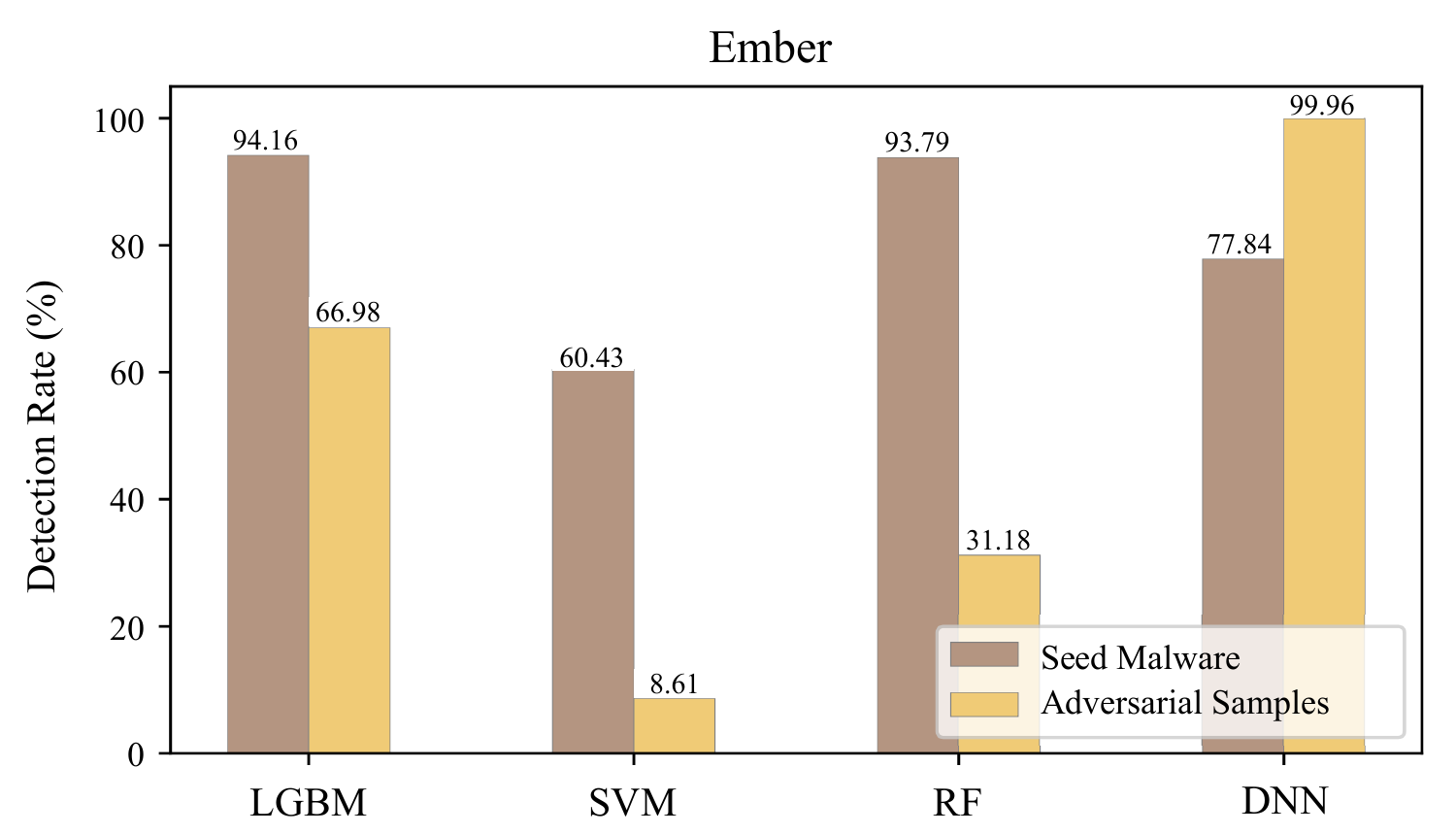}
\captionof{figure}{Detection rates of four detectors on WinPE seed malware and AMM-based test cases.} 
\label{fig_pe_result}
\end{minipage}
\hfill
\begin{minipage}[t]{.35\linewidth}
\includegraphics[width=1\linewidth]{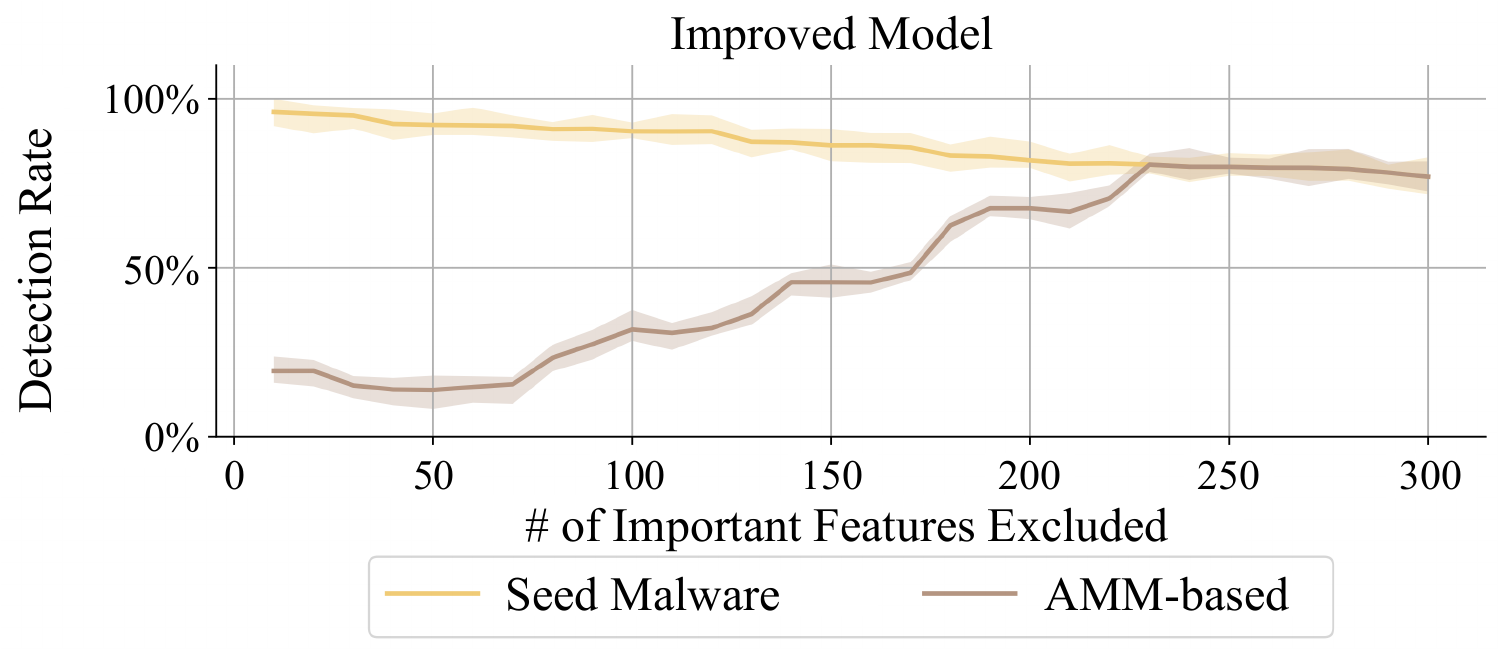}
\captionof{figure}{Detection rates of seed malware and AMM-based test cases detected by Drebin-based detectors excluding different amount of important features.} 
\label{fig_robust_param}
\end{minipage}
\vspace{-2mm}
\end{figure*}

\noindent\textbf{Feature overlap.} 
To explore the reason why the test cases can transfer across detectors, we present the top 1,024 features that have the highest AMM values in the generation models across each detector in a heatmap, shown in Figure~\ref{fig_shap_heatmap}.
In each subplot, we present the features as 32 rows by 32 columns of dots (normalized to $[0,1]$), where the darker dots represent higher values of AMM (which indicates a greater possibility to be selected as a feature to be manipulated). Further, we sort the features in the generation models according to the AMM values in descending order. Therefore, more darker dots scattered in the upper zone of the nine subplots indicate that there are more features having been selected across detectors. 
From the heatmap we can observe that 
\one features with large AMM values in LGBM (dark dots) overlap with most of the counterparts of SVM; 
and
\two many features with large AMM values in RF are out of the scope of the counterparts of LGBM. We further explain such overlaps with the result of detection rates across different models.

\noindent\textbf{Detection rate.}
We further evaluate the transferability of test cases by inspecting the detection rates.
Figure~\ref{fig_transferability_heatmap} shows the detection rates of test cases across machine learning detectors.
The y-axis represents three generation models, and the x-axis shows the target Drebin-based detectors. 
A darker color indicates a higher detection rate (representing lower transferability). 

We use the Cosine similarity (numbers in brackets) of AMM values of the top 1,024 features between models to quantify the overlaps, \ie a higher Cosine similarity value indicates a heavier overlap of high-AMM features between models.
From the results in Figure~\ref{fig_transferability_heatmap}, we observe that a higher similarity value reflects stronger transferability -- for example, the transferability from LGBM to SVM outperforms the transferability from LGBM to RF -- the overlaps resonate with the detection rate results. 
Thus, the overlaps explain why the test cases transfers across learning-based detectors. Simply put, if we manipulate enough features across different learning-based models (\ie feature overlaps), the evasion can be transferred. 

\vspace{1mm}\begin{mdframed}[backgroundcolor=black!10,rightline=false,leftline=false,topline=false,bottomline=false,roundcorner=2mm,everyline=true] 
\textbf{Takeaway 2:}  
.,The transferability of test cases depends on the overlaps of features with large Accrued Malicious Magnitude (AMM) values between different learning-based detectors.
\end{mdframed}

\subsection{Generalizability Analysis}
\label{sec_generalizability_analysis}

To examine whether our testing framework can generalize to other operating systems, we next conduct testing on Windows Portable Executable (WinPE) files. 
We use SOREL-20M~\cite{harang2020sorel20m} as the WinPE dataset in our experiment. SOREL-20M is a representative public dataset of malicious and benign WinPE samples used for malware classification, consisting of 2,381-dimensional feature vectors extracted from 9,470,626 benign and 9,919,251 malicious samples, as well as corresponding malicious binaries. 
It leverages the feature extraction function from Ember~\cite{anderson2018ember}. 
We randomly choose 10,000 benign and 10,000 malicious samples to train detectors with LGBM, SVM, RF, and DNN models. 

\noindent \textbf{Feature manipulation.~} \new{In a WinPE file, many of the features are derived from the same underlying structures, potentially leading to contradictory that cannot be manipulated concurrently. For instance, each addition of a writable section, the overall section count inevitably rises.} Previous work~\cite{alinabackdoor} shows that only 17 features can be modified directly and indirectly to preserve the functionality of WinPE binaries. The pilot experiment on WinPE dataset suggests that we should choose $N$ as 17. We leverage LIEF~\cite{LIEF} to extract features, and pefile~\cite{pefile} to apply feature manipulation on the WinPE binaries. 

\noindent \textbf{Testing results.~} 
Figure~\ref{fig_pe_result} shows the detection rates of 4 Ember-based detectors on original malware samples and test cases. 
The test cases have remarkable evasion performance on the LGBM, SVM, and RF detectors: On average, the detection rates are decreased to 
\detectionRateDecreaseAvgWinPE across the three detectors.
However, since most WinPE features correlate with each other, the method of parsing and generating WinPE binaries (\ie directly modifying values and adding empty sections) may negatively affect the evasion, illustrated by the DNN. In a nutshell, the test result shows that our proposed explainability-guided testing framework is also effective at identifying limitations in Windows malware detectors.

\subsection{Revisiting AMM with Improved Detectors}\label{sec_revisiting_amm_with_improved_detectors}

\begin{figure}[t]
\centering
\includegraphics[width=0.8\linewidth]{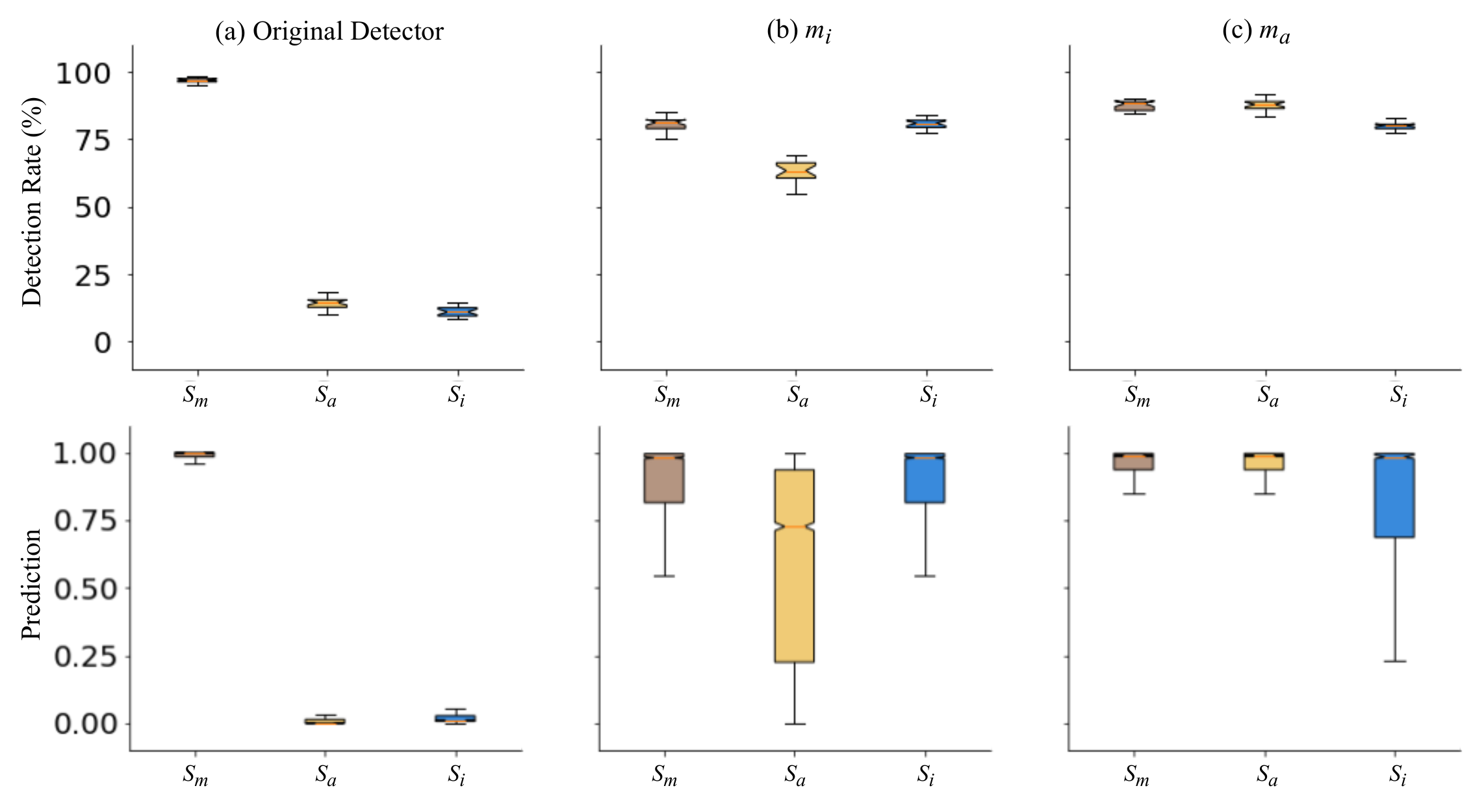}
\caption{Detection rates and prediction values of the original model and two improved models.}
\label{fig_abl_study}
\vspace{-2mm}
\end{figure}

Our evaluation shows that machine learning-based detectors are vulnerable to test cases generated by AMM. Therefore, we seek an interpretive approach to improve the robustness of existing detectors.

\noindent \textbf{Improving detectors.~}
We first follow the methodology in \S\ref{sec:malware_evaluation} and generate an improved Drebin-LGBM detector $m_i$ by excluding SAGE-based important features in the training phase. 
Specifically, we first generate new machine learning detectors each with a different number of features excluded from the training set. 
The trend of detection rates on seed malware and AMM-based test cases are shown in Figure~\ref{fig_robust_param}. 
From the result, we find that excluding 220 important features allows the improved detector to attain the highest detection rate (80.56\%) on test cases.
We therefore employ excluding 220 important features in the following experiment. 
We also generate another improved detector $m_a$ by excluding \emph{fragile} features. These are features that can be manipulated to flip the prediction, \ie features with high AMM values but low SAGE values, taking 0.2 as a threshold (the minimal AMM value of 75 features we selected in the pilot experiment is 0.2). 

\noindent\textbf{Measuring improved detectors.}~
We further generate two groups of test cases, $S_i$ (by SAGE) and $S_a$ (by AMM). 
Meanwhile we introduce the seed samples $S_m$ and the original Drebin-LGBM detector as a benchmark. In the experiment, we leverage 5,459 seed malware to generate test cases where we randomly select 300 samples for 20-round tests on each detector.
Figure~\ref{fig_abl_study} presents the detection rates and prediction values. 
We present results for three malware datasets: seed malware set $S_m$, test case sets $S_a$ and $S_i$. We define a sample as malicious when the prediction value is greater than or equals to 0.5. 
On the original detector, less than 15\% of samples in both $S_a$ and $S_i$ are classified as malicious, indicating a poor robustness against test cases. On $m_i$, the average detection rate of the $S_a$ samples increase to 67.3\% while their prediction values range from 0.23 to 0.82; in contrast, the detection rates of $S_i$ on $m_a$ are around 78.2\%, while their prediction values range from 0.69 to 0.91, which means that $m_a$ can detect more test cases than $m_i$. Note, the detection rate of $m_a$ on seed malware has been improved to 87.78\%, compared with the 80.56\% detection rate of $m_i$. 
From the experiment result, a trade-off to apply AMM on the improvement of anti-malware detector is presented: compared with the original detector, 7.34\% detection rate degradation on seed malware \vs more than 55\% increasing on the detection rate against test cases. 
This result further indicates that models with important features removed ($m_i$) is inefficient against test cases with AMM-based features manipulated ($S_a$).
Meanwhile, improved detectors guided by AMM values ($m_a$) have better robustness on seed malware and adversarial detection. 

\begin{figure}[t]
\centering
\includegraphics[width=0.8\linewidth]{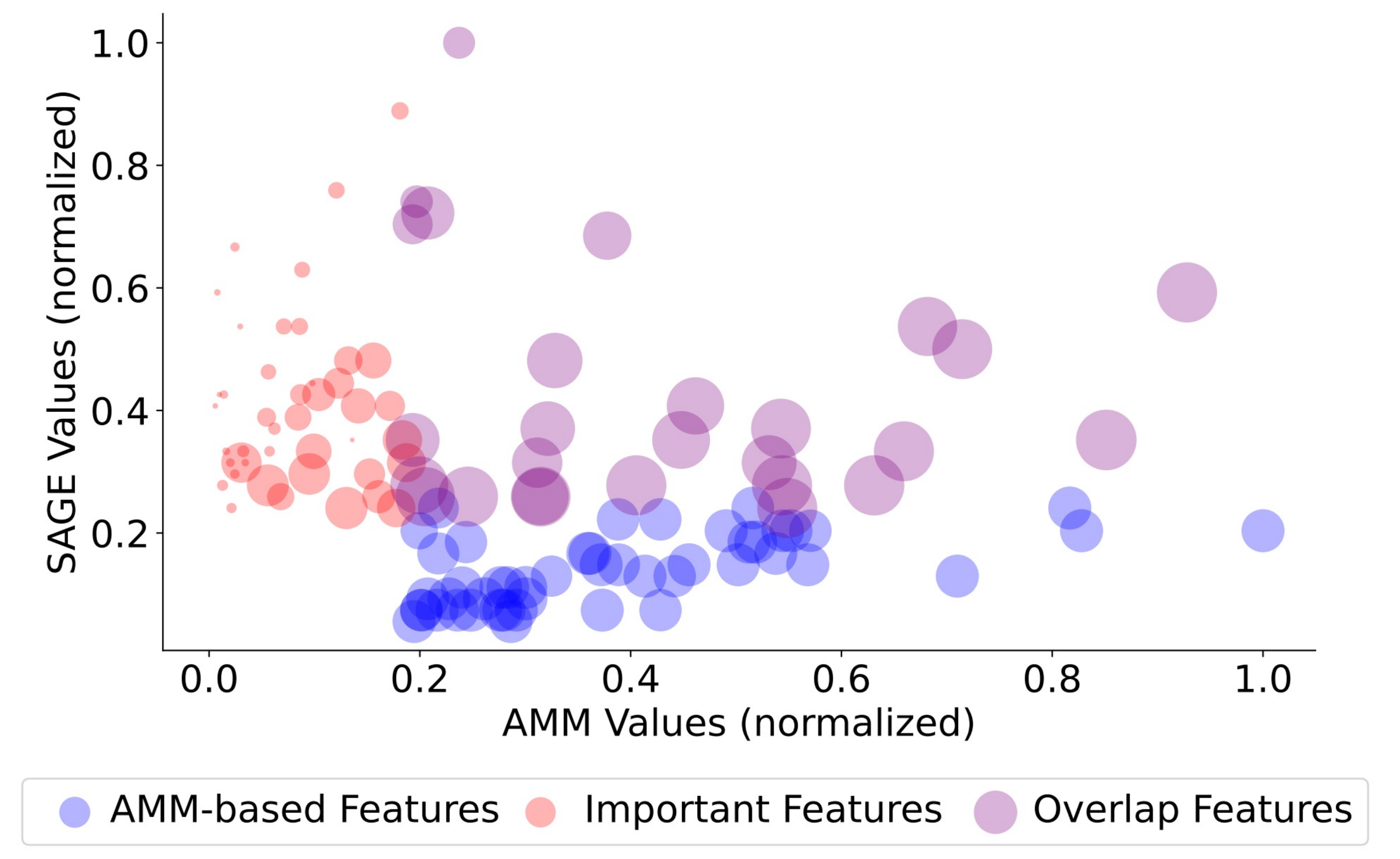}
\caption{Distributions of evaded test cases.}
\label{fig_scatter}
\end{figure}

\noindent \textbf{Comparison of AMM and SAGE.}~
Next, we compare how fragile and important features impact the detection. Figure~\ref{fig_scatter} illustrates the  distribution of top-75 AMM and SAGE featues (26 features are selected both by AMM and SAGE).  
The x-axis indicates normalized AMM values; the y-axis is normalized SAGE values; the size of the scatter point indicates the number of evaded samples manipulating the corresponding features.
As shown in the figure, samples that manipulate important features alone account for only a small portion of the evaded samples while most evaded samples are generated by manipulating features with large AMM values (pink \vs blue),
which indicates that AMM is more efficient than SAGE to generate test cases. 
In addition, this result also explains why AMM-based test cases can bypass the improved model, even they removed important features, since these models are still vulnerable to test cases that only manipulate fragile features with large AMM values. 

\vspace{1mm}\begin{mdframed}[backgroundcolor=black!10,rightline=false,leftline=false,topline=false,bottomline=false,roundcorner=2mm,everyline=true] 
\textbf{Takeaway 3:}  
\begin{itemize}[leftmargin=*]
\item Machine learning-based antivirus products should consider improving the detectors with our AMM approach.
\item AMM selects the fragile features and values that are efficient to flip the prediction results.
\end{itemize}
\end{mdframed}

\section{Discussion}\label{sec_case_studies}
In this section, we conduct two case studies to understand why some test cases can or cannot evade detection, and further discuss threats to validity of our testing framework.

\begin{figure}[t]
\centering
\includegraphics[width=0.9\linewidth]{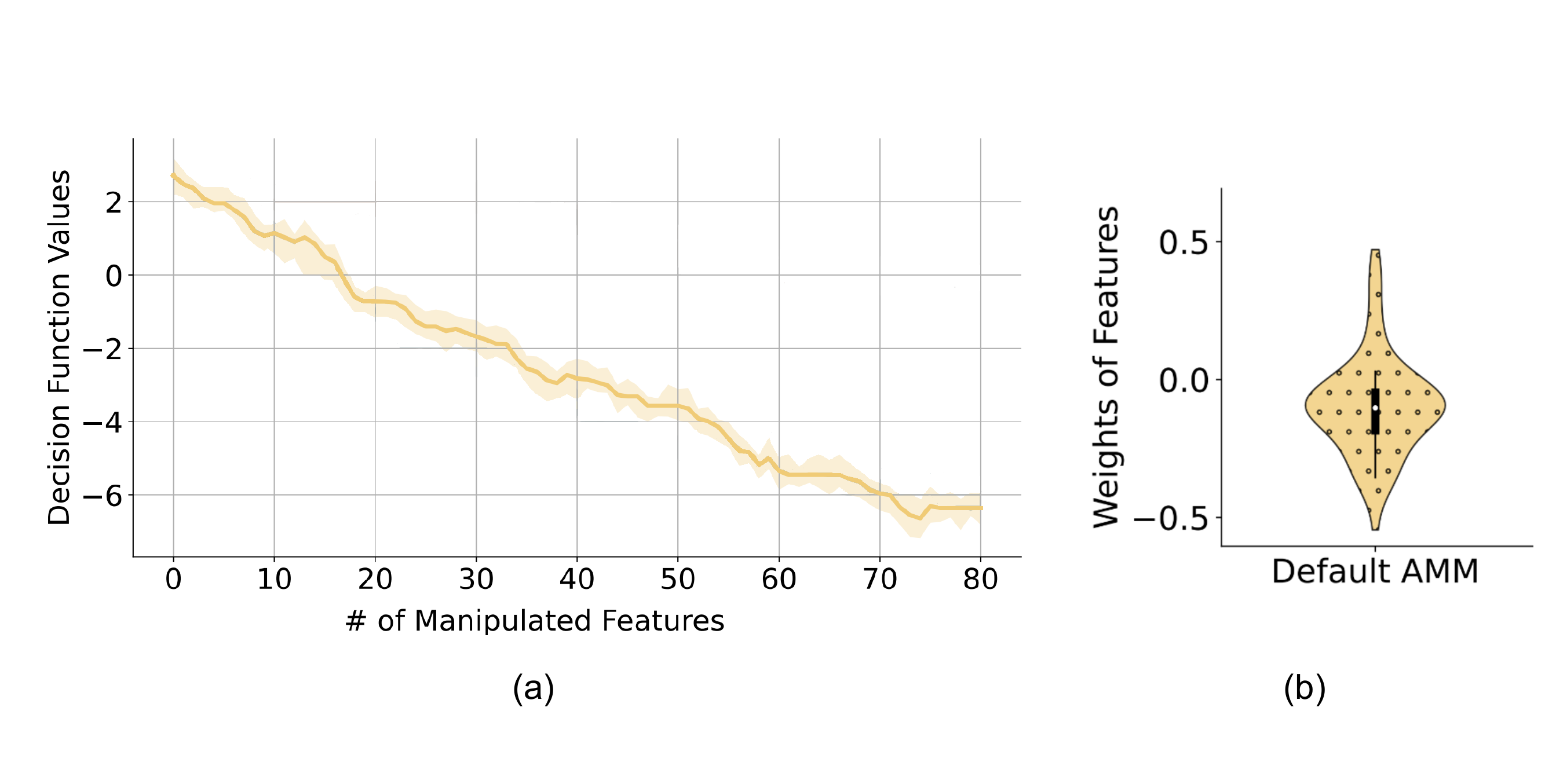}
\caption{(a) Decision function values of samples with different numbers of features manipulated; (b) Distribution of weights of features in SVM model.}%
\label{fig_case_2_df}
\end{figure}

\subsection{A Case Study on SVM Robustness}
In prior experiments, the Drebin-SVM detector could not detect test cases from all three generation strategies. Here we conduct a case study to explore the detection robustness of SVM against test cases.
First, we generate test cases by manipulating between 1 and 80 features (selected from Drebin-LGBM) on one seed. We then calculate the decision function value for each test cases, which represents the distance from the test case to the decision boundary. Positive decision function values represent malicious while negative ones represent benign.
As shown in Figure~\ref{fig_case_2_df}(a) test cases generated by AMM-based strategy invert the prediction results after manipulating 16 features and push the generation towards benign with more features manipulated. 
From the result, we speculate that the features selected from Drebin-LGBM may occupy large weights in the Drebin-SVM, so that they can invert the prediction quickly. To verify this, we export the weights of each feature in the SVM detector and compare the weight values of selected features. Figure~\ref{fig_case_2_df}(b) illustrates the weight values of the selected features, where the y-axis represents the weight of each feature. We see that most weight values of features selected by the AMM strategy from Drebin-LGBM have relatively large negative values in the SVM model, making the prediction decision values negative (\ie benign), which matches the result in Figure~\ref{fig_shap_heatmap}.
This case study confirms that features selected from LGBM occupy large negative weights in the SVM, making the prediction result of the adversarial sample benign. 

\begin{figure*}[t]
\centering
\includegraphics[width=0.85\linewidth]{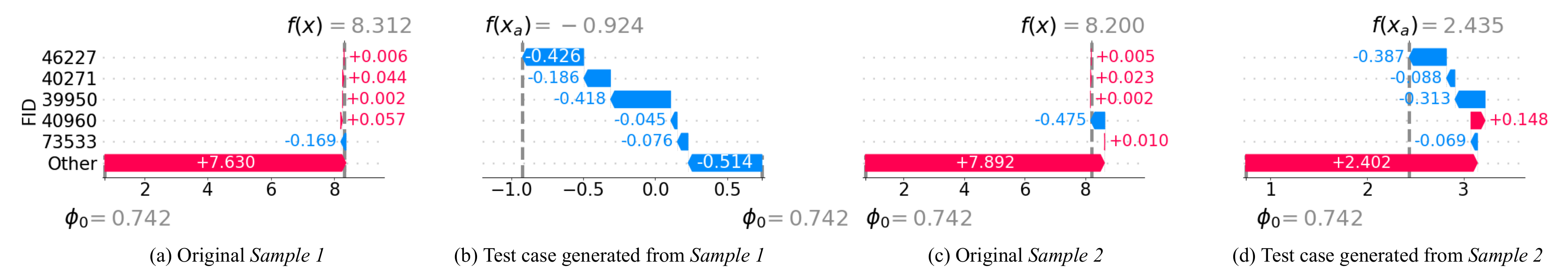}
\vspace{1mm}
\caption{SHAP values of two APK samples before and after test case generation.} %
\label{fig_shap_bar}
\vspace{-1mm}
\end{figure*}

\subsection{A Case Study on Seed Selection}

Our prior results have shown that not all test cases can evade detection. The reason could be either that the number of manipulable features in a seed is not enough to invert the prediction, or that the manipulated features have a limited impact on the prediction. To explore the reason, we choose two seed malicious APK examples, \textit{Sample~1} and \textit{Sample~2}, to generate their test cases. 
We further generate SHAP values of the original and test cases (with $N$ = 75 features selected) of \textit{Sample~1} and \textit{Sample~2} to analyze the impact of the manipulated features.

The x-axis in Figure~\ref{fig_shap_bar} indicates the prediction value, and the y-axis is the feature ID.
$f(x)$ and $f(x_a)$ are raw scores of the original and test cases given by the Drebin-LGBM detector.
We use red bars to indicate positive SHAP values, and blue bars to highlight negative SHAP values of each feature (\ie $\phi_j$ in Equation~\ref{equ:shap}). 
Feature IDs shown in the figure are parts of manipulated features that had the greatest impact on the two samples.
The test case of \textit{Sample~1} inverts its prediction as benign, and that of \textit{Sample~2} remains malicious.  Figure~\ref{fig_shap_bar}(b) shows that the SHAP values of manipulated features change significantly towards negative (blue bars), thereby pushing the output, $f(x_a)$, towards negative.
In contrast, the SHAP values of the features of \textit{Sample~2}, shown in Figure~\ref{fig_shap_bar}(d), change far less.
Specifically, only features 46227, 40271, 39950 and 73533 are manipulated towards negative with less magnitude comparing to \textit{Sample~1}, while 40960 is not towards negative at all.
This means that we cannot manipulate enough features to force the decision making towards benign for \textit{Sample~2}.
This result indicates that the manipulated features have limited impact on \textit{Sample~2} to invert the result from malicious to benign. In practice, after we increasing the number of manipulated features to 290, \textit{Sample~2} is identified as benign. 
Therefore, the capacity of a seed depends on how many features have malicious-oriented values that we can manipulate in the sample. However, infinitely increasing the number of selected features would lead to a heavy computational load and decrease the efficiency of measurement.

\subsection{Threats to Validity}

\noindent\textbf{Adaptive attacks against our testing framework.~}
Existing technology against adversarial generation, for example, adversarial training~\cite{adversarial_training} and differential privacy (DP)~\cite{abadi2016deep, dp, dfforrobust_oakland19}, could be effective against our proposed testing framework. 
For adversarial training, a machine learning-based detector could be trained with AMM generated test cases to falsely raise the detection rate on test cases. However, this may also increase the false positive rate on benign samples as the model will be forced to learn benign features as malicious, leading to a degraded performance. 
On the other hand, DP-based robust machine learning techniques cannot bypass our tests, because unbounded random perturbations may break the generated samples' functionality. 

\noindent\textbf{Dynamic detection.~}
Since we only insert static unreachable instructions into the malware sample, a detector with dynamic detection will easily pass our testing. Feature-space manipulation and problem-space obfuscation rely on static syntactic and structural modification. These modifications can be used to test static machine learning-based detectors and rule-based antivirus engines. However, the malicious behaviors preserved in the test cases will still be exposed during run-time and identified by the detectors that adopt dynamic analysis. 
Considering that dynamic feature detection consumes more resources to monitor, this approach may be impractical on a large scale and static approaches are still widely used in practice. In this research, we focus on exposing the weaknesses of malware detectors with an explainable method.

\section{Conclusion}

This paper has proposed an explainability-guided malware detector testing framework.
The framework performs test case generation, relying on Accrued Malicious Magnitude (AMM) to guide the feature selection and a binary builder to map feature-space manipulations onto problem-space binaries. 
We then use our framework to test the robustness of state-of-the-art malware detectors.
Our research includes the following key findings: 
\one commercial antivirus engines and state-of-the-art machine learning detectors are vulnerable to AMM-based test cases; 
\two the transferability of AMM test cases relies on the overlaps of features with large AMM values between different machine learning models;
and
\three AMM values can effectively measure the fragility of features and explain the capability of flipping classification results. 
According to our findings, we suggest that machine learning-based AV products should consider using the AMM values to improve their robustness. 
Exploring the latter constitutes our key line of future work, as we believe this could prompt a new approach to defending against malware evasion attacks. 

\section*{Data Availability}
The source code, excluding the test case generator, of this project is publicly available at \url{https://github.com/Immor278/AMM}. Considering the potential security issues, we will not release the test case generator and any test cases, as well as the information of commercial antivirus involved in our evaluation, except for academic uses that are approved by our institutional ethics committee.

\section*{Acknowledgments}
The work has been supported by the Cyber Security Research Centre Limited whose activities are partially funded by the Australian Government’s Cooperative Research Centres Program. Minhui Xue and Ruoxi Sun are the corresponding authors of this paper. 
\balance
\bibliographystyle{ACM-Reference-Format}
\bibliography{ref}

\end{document}